\documentclass[11pt]{iopart}

\usepackage{overpic}
\usepackage{xfrac}
\usepackage{nicefrac}
 \expandafter\let\csname equation*\endcsname\relax
 \expandafter\let\csname endequation*\endcsname\relax
\usepackage{amsmath}
\usepackage{mathtools}
\usepackage{tabularx}
\usepackage{color}
\definecolor{ao}{rgb}{0.0, 0.5, 0.0}

\usepackage[colorlinks=true,
 linkcolor=black,
 urlcolor=blue,
 citecolor=blue]{hyperref}

\newcommand{\E}{\mathcal{E}}

\renewcommand{\vec}[1]{\mathbf{#1}}

\renewcommand{\vec}[1]{\mbox{\boldmath$#1$}}

\newcommand{\beq}{\begin{equation}}
\newcommand{\eeq}{\end{equation}}
\newcommand{\bal}{\begin{aligned}}
\newcommand{\eal}{\end{aligned}}

\newcommand{\bd}{\begin{displaymath}}
\newcommand{\ed}{\end{displaymath}}

\newcommand{\erf}{{\rm erf}}

\newcommand{\BE}{\begin{eqnarray}}
\newcommand{\EE}{\end{eqnarray}}
\newcommand{\be}{\begin{equation}}
\newcommand{\ee}{\end{equation}}

\newcommand{\NTCP}{\mbox{NTCP}}

\def\d{\mathrm{d}}

\def\E{\mathcal{E}}

\def\half{\nicefrac{1}{2}}
\def\nhalf{-\nicefrac{1}{2}}

\begin{document}

\title[Calculating normal tissue complication probabilities from stochastic ...]{Calculating normal tissue complication probabilities and probabilities of complication-free tumour control from stochastic models of population dynamics}

\author{Peter G. Hufton}

\address{Theoretical Physics, School of Physics and Astronomy, The University of Manchester, Manchester M13 9PL, United Kingdom}
\ead{peter.hufton@postgrad.manchester.ac.uk}

\author{Elizabeth Buckingham-Jeffery}
\address{School of Mathematics, The University of Manchester, Manchester M13 9PL, United Kingdom}
\ead{e.buckingham-jeffery@manchester.ac.uk}

\author{Tobias Galla}

\address{Theoretical Physics, School of Physics and Astronomy, The University of Manchester, Manchester M13 9PL, United Kingdom}
\ead{tobias.galla@manchester.ac.uk}
\vspace{10pt}
 
\begin{abstract}
We use a stochastic birth-death model for a population of cells to estimate the normal tissue complication probability (NTCP) under a particular radiotherapy protocol. We specifically allow for interaction between cells, via a nonlinear logistic growth model.
To capture some of the effects of intrinsic noise in the population we develop several approximations of NTCP, using Kramers--Moyal expansion techniques. These approaches provide an approximation to the first and second moments of a general first-passage time problem in the limit of large, but finite populations. We use this method to study NTCP in a simple model of normal cells and in a model of normal and damaged cells. We also study a combined model of normal tissue cells and tumour cells. Based on existing methods to calculate tumour control probabilities, and our procedure to approximate NTCP, we estimate the probability of complication free tumour control. 
\end{abstract}

%
%
%
%
%

\section{Introduction}

When giving a dose of radiation to a tumour it is likely that the surrounding healthy tissue will also be damaged. 
A radiotherapy treatment protocol aims to provide enough radiation to the tumour to control the cancer whilst not causing excessive side-effects by damaging surrounding tissue. 
To this end, a protocol must find a balance between maximising the tumour control probability (TCP) and minimising the normal tissue complication probability (NTCP).
Normal tissue complications (NTCs) encompass a wide variety of problems ranging in severity from increased urinary frequency from the treatment of prostate cancers \cite{martinez2015permanent,tanaka2015urethral} to severe neurological complications such as myelitis from the treatment of neck cancers \cite{horiot1997accelerated} and organ failure \cite{martinez2015permanent}.

There are numerous models of TCP and NTCP in the literature. 
Broadly, the term `model' is used to describe two different types of mathematical approaches to characterising these probabilities. 
The first is statistical: based on cohorts of patients statistical models are developed to identify factors contributing to the TCP and the NTCP. This is then used to find mathematical expressions which allow one to estimate the TCP or NTCP of a patient with given characteristics and for given radiation protocols \cite{Lyman1985,Niemierko1993}. 
The focus of our work is not on this type of model. 
Instead, we concentrate on the second type of modelling approach, which seeks to compute TCP and NTCP `bottom-up' from mechanistic principles of the population dynamics of tumour or normal cells \cite{hanin2013mechanistic,zaider2000tumour}. 
These models are often stylised, but the key characteristic they all share is that they describe the dynamics of cell division and death. Many of these models are intrinsically stochastic. Mitosis and cell death are random events in such models, and the precise outcome is therefore uncertain; the tumour may or may not be controlled, and NTCs can arise, but do not have to. The aim of this line of research is to obtain, for a given model of the population dynamics of cells and a given radiation protocol, the TCP and NTCP. 
The word `obtain' includes by computer simulation of the population, or by direct mathematical computation when this is possible. 
While simulations are sometimes viable, the mathematical route, when it is available, is generally preferable as explicit formulae provide an efficient way of evaluating TCP or NTCP, often much faster than simulation. Not all types of population dynamics can be treated mathematically exactly however. In such cases approximations have to be made in the mathematical calculation of TCP and NTCP.

TCP from a stochastic birth-death model has previously been described by Zaider and Minerbo~\cite{zaider2000tumour}, subsequent work includes \cite{dawson2006derivation,maler2009cell,hillen2010cell}. A stochastic birth-death model of normal tissue cells was described by Stocks \emph{et al.}~\cite{stocks2016stochastic}, but their mathematical calculation of NTCP ultimately does not take into account intrinsic stochasticity in the population. We extend this analysis and capture features of intrinsic noise in the calculation of NTCP. We use a stochastic birth-death model of normal tissue cells where cell death rates are affected by the dose and timing of radiotherapy. NTCP can be seen as the cumulative distribution function of the first-passage time of this stochastic birth-death process through a boundary; NTC sets in when the number of functional cells falls below a certain threshold. We obtain estimates of NTCP by approximating the distribution of first-passage times. 

One may ask whether the inclusion of intrinsic noise is necessary in modelling NTCP. Hanin and Zaider \cite{hanin2013mechanistic} argue that deterministic approaches might be sufficient, due to the high numbers of cells involved. However we note that the size of the population may vary depending on context. For example, the model could describe a functional subunit (FSU) of an organ, rather than the entire organ \cite{Niemierko1993,stavrev2001generalization,tucker2006cluster}. NTCP would then not necessarily indicate the probability that an organ fails, but instead that such a subunit no longer fulfils its function. For instance, Niemierko and Goitein consider a kidney split into $10^7$ FSUs, where each FSU contains $10^4$ cells \cite{Niemierko1993}. In such circumstances noise in the population (i.e., within a FSU) may become relevant. Intrinsic stochasticity may also be important in the context of stem cells, especially if they are present in relatively small numbers \cite{Rutkowska2010,dandrea2016,hendry1986tissue,konings2005mechanism}. It is also interesting to note that some of the statistical models mentioned above assume a normal distribution of NTC onset, see e.g., the model proposed by Lyman in Ref.~\cite{Lyman1985}. The resulting NTCP then takes the form of an error function, i.e., the integral of a Gaussian distribution, similar to what we find from our approximations. It is important to note though that the origin of stochasticity may be different, as discussed in more detail in our conclusions.

Mathematically, our main result is intuitive. We find that, for a sufficiently large population, the distribution of first-passage times through the threshold at which an NTC sets in is approximately normal. The variance of this normal distribution decreases proportionally to the size of the population. The deterministic result for NTCP by Stocks \emph{et al.}~\cite{stocks2016stochastic} is recovered in the limit of infinite population size (NTCP as approximated by Stocks \emph{et al.} was either zero or one).

While our approximation is relatively crude, the mathematical simplicity of our result is a strength. Using our method to predict NTCP does not require extensive numerical calculations. In some examples closed-form expressions can be obtained, in other cases a small set of ordinary differential equations (ODEs) needs to be solved numerically, which can be done much more efficiently than integrating forward a potentially high-dimensional master equation. Since the linear-noise approximation (LNA), on which our approach is based, is ubiquitous in statistical physics and applications, our result may also lend itself to applications in other fields outside of radiotherapy modelling. 

The remainder of this paper is set out as follows. In Sec.~\ref{sec:model} we present the microscopic model of normal tissue cells adapted from the model of Stocks \emph{et al.}~\cite{stocks2016stochastic} and a definition of NTCP. We use this model to explain the steps of our approximation and derive our main results. This involves first writing the master equation, and subsequently approximating the dynamics by carrying out a Kramers--Moyal expansion and LNA. We then proceed to approximate the first-passage time across a boundary by considering the dynamics in a small region near the boundary marking the onset of NTC. This provides a Gaussian approximation of the first-passage times, and thus an approximation to NTCP. Following Hanin and Zaider~\cite{hanin2013mechanistic} we then consider a more complicated model of normal tissue in Sec.~\ref{sec:2d}. In this model there are two types of cells (normal and damaged), and we show how our method can be extended to systems with more than one degree of freedom. In the context of this model we also develop a second approximation method for NTCP. In Sec.~\ref{sec:CFC} we combine models of cancerous cells and normal tissue to estimate the probability of complication-free tumour control, i.e., the probability that the tumour is controlled without complications in the normal tissue. In Sec.~\ref{sec:concl} we finally summarise our results. The Appendix contains further details of our analysis.

\section{Logistic model of healthy tissue}
\label{sec:model}
\subsection{Model definitions}\label{ssec:model}
We first focus on a model of normal tissue similar to that in Ref.~\cite{stocks2016stochastic}, which is itself an individual-based extension to the deterministic dynamics considered in Ref.~\cite{hanin2013mechanistic}. This existing work produced analytical descriptions of NTCP, but the analysis was restricted to the deterministic limit, in which intrinsic noise within the population is discarded. Our approach retains some of the effects of demographic noise on NTCP.

The model describes a well-mixed population of cells, we write $N_t$ for the size of the population at time $t$. Cells can divide by mitosis at a rate $b$. We assume that overall growth is limited by spatial constraints and the presence of nutrients, so that $b$ is a logistic function of $N$,
\begin{equation}
b_N= \left\{ \begin{array}{cl}
 b_0\left(1 - \tfrac{N}{K} \right) &\mbox{if } N\leq K \\
0 & \mbox{otherwise},
\end{array} \right.
\end{equation}
where $b_0>0$ is a constant parameter. This indicates that the per capita birth rate decreases with increasing population size, and growth ceases completely when the carrying capacity $K$ is reached; $K$ is a model parameter and constant in time.

Cells can die due to natural causes and from external radiation. Natural death occurs with rate $d$. We note that explicitly separating death processes from birth events is necessary for a stochastic treatment of the model; basing the analysis on an effective net growth rate (i.e., $b_N-d$), as in Ref.~\cite{stocks2016stochastic}, is insufficient to model the dynamics outside of the deterministic limit (models with different birth and death rates, but with the same net growth rate can lead to different results for NTCP in a stochastic setting).

External radiation damages cells mainly by inducing single or double strand breaks in their DNA \cite{dale2007radiobiological}. The model captures these processes via a hazard function $h(t)$, denoting the per capita death rate due to radiation. This rate will generally depend on time, as determined by the details of the applied radiation protocol. For example, we consider the linear-quadratic (LQ) formalism of brachytherapy in Sec.~\ref{sec:2d}.

The model can be summarised as a list of `reactions', with notation similar to that used in chemical reaction systems. We write $\mathcal{N}$ to represent an individual normal cell. The dynamics are then given by
\begin{align}
\begin{aligned}[c]
\mathcal{N} \xrightarrow{\mathmakebox[15mm]{ b_0 \left(1 - \tfrac{N}{K} \right) }}{}& \mathcal{N}+\mathcal{N}~~&&\text{(mitosis)},\\
\mathcal{N} \xrightarrow{\mathmakebox[15mm]{ d }}{}& \emptyset &&\text{(natural death)}, \\
\mathcal{N} \xrightarrow{\mathmakebox[15mm]{ h(t) }}{}& \emptyset &&\text{(death due to radiation)},
\end{aligned}
\label{eq:reactions_simple}\end{align}
where the rates above the arrows are per capita rates. 

The deterministic rate equation for this system can be formulated heuristically as follows,
\be\label{eq:det}
\frac{\d N}{\d t}=b_0 N \left(1 - \tfrac{N}{K} \right)-[d+h(t)]N.
\ee
It can also be derived systematically from the lowest-order terms in an expansion in the inverse system size, as discussed below.

In the absence of radiation [i.e., when $h(t)=0$], the non-zero fixed point of Eq.~\eqref{eq:det} is given by $N^*=K\left(1-\frac{d}{b_0}\right)$.
Since the population dynamics are stochastic, the size of the population fluctuates about this value. To simplify the notation we will use $K=\frac{M}{1-d/b_0}$ in the following, such that---in the absence of radiation---the average population size is $M$.

\subsection{Master equation}
The process defined by Eqs.~\eqref{eq:reactions_simple} can equivalently be described by a (chemical) master equation (CME). This is a set of ODEs describing the evolution in time of the probability for the population to be in each of the possible states, $N$. We write $P_N(t)$ for the probability that the population has size $N$ at time $t$. The master equation is then given by
\begin{align}
\begin{split}\frac{\d}{\d t} P_N(t) =
~&\left(\E^{-1}-1\right) N b_0 \left(1 - \tfrac{N}{K} \right)P_N(t)\\
+&\left(\E-1\right) N \left[d+h\left(t\right)\right] P_N(t),\end{split}
\label{eq:CME}
\end{align}
where $\E$ is the step operator defined by its effect on a function $f_N$, i.e., we have $\E f_N=f_{N+1}$, and similarly, $\E^{-1} f_N=f_{N-1}$. The operators act on everything to their right.

\subsection{Definition of normal-tissue complication probability and strategies to calculate it}
\subsubsection{Definition}

An organ requires a minimum number of cells to function properly \cite{bond1965mammalian}. We introduce a threshold, $L$, and say that a normal tissue complication (NTC) is encountered when the number of cells in the population $N_t$ falls below $L$. Given that $N_t$ is a stochastic process, NTC will occur at different times in different realisations of the model dynamics (or potentially, it may never occur in a given realisation). This leads to the definition of normal tissue complication probability (NTCP). We assume that once NTC has been encountered in a given realisation of the dynamics, it cannot be repaired, even if the number of cells ultimately recovers to values above the threshold $L$. We therefore define $\NTCP(t)$ as the probability that, at some time before $t$, the population contained $L$ cells or fewer. NTCP is then by definition an increasing function of time. We remark that this definition of NTCP$(t)$ differs from one used previously in Ref.~\cite{stocks2016stochastic}, which allowed NTCP$(t)$ to decrease. In practice results using the two different definitions are often very similar.

Mathematically the calculation of NTCP constitutes a first-passage time problem \cite{redner2001guide}. More precisely, NTCP$(t)$ is the cumulative distribution function of the first-passage time through the threshold $L$. The methods we develop to approximate NTCP are therefore potentially applicable to a variety of other problems involving the estimation of first-passage time distributions, beyond the specific example of NTCP.

\subsubsection{Strategies for the calculation or simulation of NTCP}

Realisations of the process defined by Eqs.~\eqref{eq:reactions_simple} can be generated using the stochastic simulation algorithm by Gillespie \cite{gillespie1976general,gillespie1977exact}. In principle, a large ensemble of such simulations can be used to measure NTCP$(t)$. However, in practice this approach is of limited use since a large number of runs need to be collected to obtain sufficient statistics. Simulations also offer relatively little in the way of mechanistic insight.

One can also find the $\NTCP(t)$ by direct numerical integration of Eq.~\eqref{eq:CME}. To do so, one must impose an absorbing boundary at $L$, i.e., the birth rate $b_L$ would have to be set to zero so that once a trajectory has reached the threshold $L$ it cannot recover to values above the threshold.
In practice, this approach is computationally costly, especially in more realistic models where there are several different types of cells (see e.g., Sec.~\ref{sec:2d}). The master equation is then a large set of coupled ODEs which would have to be integrated forward. 

An alternative approach involves the use of generating functions (for general principles see for example Ref.~\cite{gardiner1985handbook}). However, this technique is usually only viable for relatively simple models. For example, generating functions can sometimes be calculated analytically when per capita birth and death rates do not depend on the current population size, i.e., when $b_N$ is independent of $N$. This indicates that different cells reproduce and die independently of each other, and for such models explicit equations for both TCP and NTCP can, in principle, be obtained based on generating functions. This is not the case in the above logistic growth process however, which involves interaction between cells due to the overall carrying capacity. A notable example of an exact calculation using generating functions is the work of Zaider and Minerbo in Ref.~\cite{zaider2000tumour} who obtain TCP in closed form for a linear-birth death process with time-dependent death rate (the time dependence is due to irradiation of the population). Their result for TCP can be expressed in terms of the solution of the rate equation describing the population in the deterministic limit (see also Ref.~\cite{gong2011}). It is important to note though the result of Ref.~\cite{zaider2000tumour} for TCP is valid for populations of any finite size, whereas the approximation of NTCP in Ref.~\cite{stocks2016stochastic} discards intrinsic fluctuations.

Given the limitations of these numerical and analytical methods, we develop and use an approximation to estimate the NTCP. The approach is based on Kramers--Moyal expansion techniques \cite{van1992stochastic,gardiner1985handbook} and retains features of the intrinsic noise resulting from the finiteness of the population of cells. At the same time, we assume that the population is sufficiently large so that the jump process defined by the master equation (\ref{eq:CME}) can be approximated by a stochastic differential equation (SDE). 

\subsection{Kramers--Moyal expansion and linear-noise approximation}
\subsubsection{Kramers--Moyal expansion and Fokker--Planck equation}
The expansion method is based on the assumption of a large, but finite population, as will be explained in further detail below. We will refer to $M$ as the system size, in-line with previous literature \cite{van1992stochastic,gardiner1985handbook}. As a first step we introduce the population density $n_t=N_t/M$; that is, the population size at time $t$ divided by the typical system size. We re-scale the threshold for the onset of NTC in the same way and write $\ell=L/M$; NTC thus occurs when $n_t\leq \ell$. We also introduce a re-scaled carrying capacity and write $k=K/M$. Given our above choice $K=\frac{M}{1-d/b}$, we have $k=(1-d/b)^{-1}$. 

Re-writing functions of $N$ as functions of $n=N/M$, we find $\E^{\pm 1} f(n)=f(n\pm 1/M)$ for the action of the step operator. We proceed to consider the limit where the system size is large, $M\gg1$. In this limit one can expand
\begin{equation}
\E^{\pm 1}=1\pm\frac{1}{M}\frac{\partial}{\partial n}+\frac{1}{2M^2}\frac{\partial^2}{\partial n^2}+\dots \text{ .}
\end{equation}
Substituting this into the master equation~\eqref{eq:CME} results in a Fokker--Planck equation for the probability density $\Pi(n, t)$,
\begin{align}\begin{split}
\frac{\partial}{\partial t}\Pi(n,t)=
&-\frac{\partial}{\partial n}\mu(n,t)\Pi(n,t)
+\frac{1}{2M}\frac{\partial^2}{\partial n^2} \sigma^2(n,t)\Pi(n,t),
\label{eq:FPE}
\end{split}\end{align}
where we have neglected higher-order terms in $M^{-1}$. The probability of finding the random process $n_t$ with a value in the interval $[n, n+\d n)$ at time $t$ is $\Pi(n,t)\d n$.

For the current model, the drift and diffusion terms in Eq.~(\ref{eq:FPE}) are given by
\begin{subequations} \begin{align}
\mu(n,t)&=nb_0\left(1-\frac{n}{k}\right)-n\left[d+h(t)\right], \label{eq:DriftDiffusion} \\ \sigma^2(n,t)&=nb_0\left(1-\frac{n}{k}\right)+n\left[d+h(t)\right],
\end{align} \end{subequations}
respectively. Equation~\eqref{eq:FPE} describes the statistics generated by solutions of the It\={o} SDE
\begin{align}
\d n_t = \mu(n_t,t) \d t + M^{-1/2}\sigma(n_t,t) \d W_t,
\label{eq:SDE}
\end{align}
where $W_t$ is a standard Wiener process.

In principle, trajectories of this SDE can be generated in simulations, for example using the Euler--Maruyama method~\cite{kloeden1992sde}. These simulations are more efficient than simulating the original model, in particular the population size only enters in the noise strength and does not affect computing time required to generate a set number of realisations. However, our aim is to make analytical progress. This requires further approximation, first because $\mu(n_t,t)$ is a non-linear function of $n_t$, and more importantly because the noise in Eq.~\eqref{eq:SDE} is multiplicative. We proceed by making a further simplification using the LNA \cite{gardiner1985handbook,van1992stochastic}, effectively turning multiplicative noise into additive noise.

\subsubsection{Linear-noise approximation}\label{sec:LNA}
To carry out the LNA we introduce the stochastic process $\xi_t$ via the transformation \cite{van1992stochastic}
\begin{align}
n_t = \phi(t) + M^{-1/2}\xi_t,
\label{eq:LNA_ansatz}
\end{align}
where $\phi(t)$ is a deterministic function of $t$, to be determined shortly.
 
We next substitute this ansatz into Eq.~\eqref{eq:SDE}, and expand in powers of $M^{-1/2}$. From the two lowest-order terms we find
\begin{subequations}\begin{align}
\frac{\d \phi}{\d t} ={}& \mu\left[\phi(t),t\right]\label{eq:LNA_top}, \\
\d \xi_t ={}& \mu'\left[\phi(t),t\right] \xi_t \d t + \sigma\left[\phi(t),t\right] \d W_t\label{eq:LNA_bottom},
\end{align}\label{eq:LNA}\end{subequations}
where $\mu'\left[\phi(t),t\right]$ is the derivative of the drift $\mu(n,t)$ with respect to $n$, evaluated at $\phi(t)$ and $t$.

The first of these equations indicates that $\phi(t)$ is the solution of a deterministic rate equation. Up to re-scaling of $N$ and $K$ this rate equation is identical to Eq.~(\ref{eq:det}). The SDE (\ref{eq:LNA_bottom}) describes fluctuations about this deterministic trajectory, due to demographic noise. We note that the LNA is only valid provided corrections to the deterministic dynamics remain small; if this is not the case higher-order terms in the system-size expansion become important. The approximation is generally appropriate if the deterministic trajectory is locally attracting, i.e., if $\mu'[\phi(t),t]<0$ at all times. This condition is fulfilled in the present model.

The linear SDE (\ref{eq:LNA_bottom}) can be solved straightforwardly \cite{van1992stochastic,gardiner1985handbook,risken1984fokker}, and, within the LNA, the distribution of $n_t$ is found to be Gaussian, centred around the solution $\phi(t)$ of Eq.~(\ref{eq:LNA_top}),
\begin{align}
\Pi(n,t)
=&\frac{1}{\sqrt{2\pi M^{-1}\Sigma^2(t)}}\exp\left(-\frac{[n-\phi(t)]^2}{2M^{-1}\Sigma^2(t)}\right).
\label{eq:dist}
\end{align}

The variance of this distribution, $M^{-1}\Sigma^2(t)$, is a function of time, and can be obtained from the solution of 
\begin{align}
\frac{\d \Sigma^2}{\d t} = 2\mu'\left[\phi(t),t\right]\Sigma^2(t) + \sigma^2\left[\phi(t),t\right],
\label{eq:variance}
\end{align}
see e.g., Ref.~\cite{risken1984fokker}.

For some cases Eqs.~\eqref{eq:LNA_top} and~\eqref{eq:variance} can be solved exactly, and one can obtain an analytical expression for $\Pi(n,t)$ in Eq.~(\ref{eq:dist}). We discuss this in the context of the current model below.
For the general case, these equations can be integrated forward numerically, using standard Runge--Kutta methods. This only requires the integration of two ODEs.

\subsubsection{Approximation of $\NTCP(t)$}\label{ssec:approx1}
\begin{figure*}[t]
\center
\includegraphics[width=\textwidth]{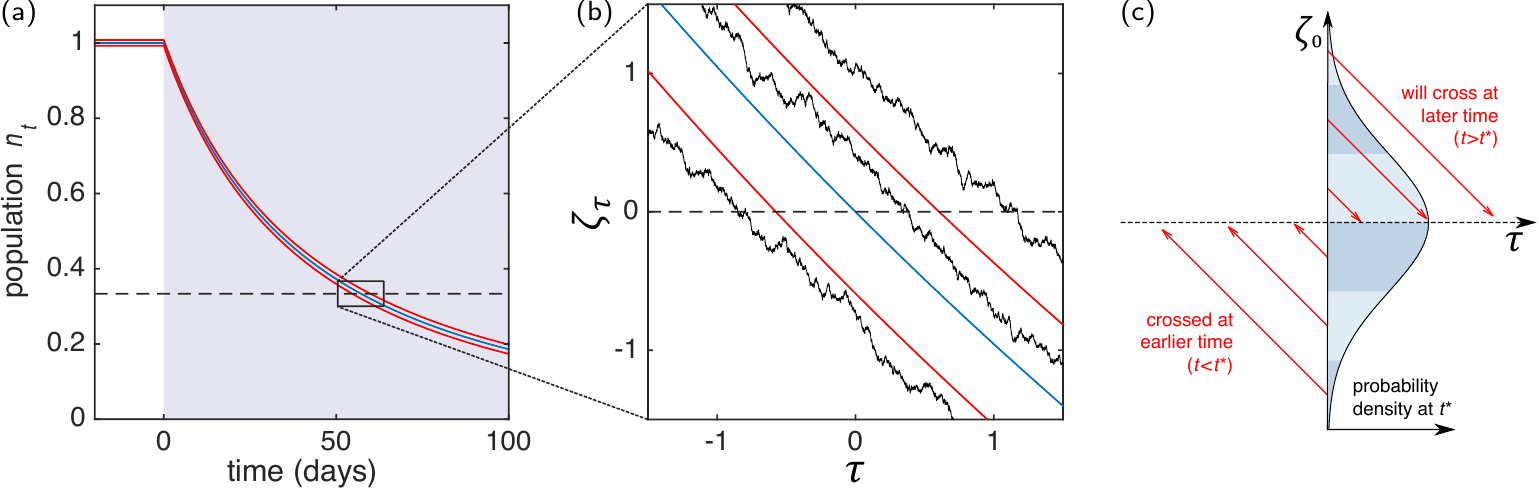}
\caption{\emph{Population size as a function of time for the model in Sec.~\ref{ssec:model}.} In this set-up constant radiation acts from a given time, here chosen to be $t=0$. The size of the population then decreases and falls below the threshold for the onset of NTCs. Panel (a): The central, blue line shows the deterministic trajectory [Eq.~(\ref{eq:LNA_top})], the red lines show a band of one standard deviation as predicted by the LNA, see Eq.~(\ref{eq:variance}). The shading of the background indicates the rate of cell death due to radiation $h(t)$. The dashed line is the threshold for onset of NTC. Panel (b): Magnified look at the crossing region, shown in the re-scaled coordinates $\tau$ and $\zeta$. Shown are three stochastic trajectories (black noisy lines) from simulation of SDE~\eqref{eq:SDE}; they are approximately linear with gradient minus one, as predicted by Eq.~\eqref{eq:linear_path}.
Panel (c): Schematic representation of our approximation. We start from the Gaussian distribution obtained within the LNA [Eq.~\eqref{eq:dist}] and project trajectories onto the time axis, assuming that their behaviour is linear with slope minus one. Model parameters are given in Table~\ref{table:parameters_simple} [parameter set (D)].
}
\label{fig:process}
\end{figure*}
We now proceed to estimate NTCP using the outcome of the LNA. Taking Eqs.~\eqref{eq:LNA_top} and \eqref{eq:LNA_bottom} as a starting point, the calculation of NTCP amounts to a first-passage time problem for a SDE with time-dependent drift and noise strength.
Equation~\eqref{eq:LNA_bottom} describes an Ornstein--Uhlenbeck process with time-dependent rates \cite{gardiner1985handbook}. Due to the time-dependence of $\phi(t)$ in Eq.~(\ref{eq:LNA_top}), calculating NTCP amounts to calculating the first-passage time of this Ornstein-Uhlenbeck process through a moving boundary. While the first-passage time distribution of Ornstein--Uhlenbeck processes is available for constant rates and a static boundary \cite{ricciardi1988first}, studies of instances with time-dependence are often based on approximation schemes for specific cases; examples can be found in Refs.~\cite{madec2004first,lo2006computing}.

To make progress we therefore use a further approximation. We focus on cases in which the deterministic trajectory $\phi(t)$ crosses the threshold $\ell=L/M$, as illustrated in Fig.~\ref{fig:process}(a); we write $t^*$ for this time. The exact value of $t^*$ will depend on the applied radiation protocol and the other model parameters. The calculation of $\NTCP(t)$ by Stocks \emph{et al.} \cite{stocks2016stochastic} is based on this deterministic contribution, and within their calculation $\NTCP(t)=\Theta(t-t^*)$ is a Heaviside step function [$\Theta(u)=1$ for $u\geq 0$, and $\Theta(u)=0$ otherwise]. Our aim is to build on the results in Ref.~\cite{stocks2016stochastic} and to capture some of the influence of intrinsic fluctuations on NTCP. 

As a next step we look at the dynamics of Eqs.~(\ref{eq:LNA_top}) and (\ref{eq:LNA_bottom}) in a time window around $t^*$, as shown in Fig.~\ref{fig:process}(b). Some trajectories of the stochastic system will cross the threshold $\ell$ before $t^*$, and others after $t^*$. We expect these fluctuations in the crossing time to decrease as the system-size parameter $M$ is increased. To evaluate this further we consider the Gaussian distribution for the population density $n_{t^*}$ obtained by evaluating Eq.~(\ref{eq:dist}) at time $t^*$. By construction, this distribution is centred on $\ell$, as shown in Fig.~\ref{fig:process}(c). We now proceed on the basis that trajectories with values $n_{t^*}>\ell$ will first cross the threshold at a time greater than $t^*$, and estimate this time of crossing from the dynamics near $t^*$. Similarly, trajectories with $n_{t^*}<\ell$ have already crossed the threshold, and we estimate how long before $t^*$ this has occurred. This procedure implies several assumptions, for example a trajectory with $n_{t^*}>\ell$ may have had its first crossing before $t^*$ and then returned to values $n_t$ above $\ell$ due to further fluctuations. This is not captured by our estimate of NTCP.

\begin{table*}[t!]
\small\centering
\begin{tabularx}{1\linewidth}{|llXl|}
 \hline
 &Coordinate &Interpretation & Relations \\ [0.5ex]
 \hline
(A)&$N_t$ & number of individuals in population at time $t$ & --- \\ \hline
(B)&$n_t$ & population density& $n_t={N_t}/{M}$ \\ \hline
(C)&$\phi(t)$ & deterministic (mean-field) trajectory & $n_t=\phi(t)+M^{\nhalf}\xi_t$ \\
 &$\xi_t$ & deviation from mean-field path due to linear noise &\\ \hline
(D)&$\zeta_\tau$ & re-scaled population near boundary $l=L/M$ & $n_\tau=\ell+M^{\nhalf} \zeta_\tau$ \\
& $\tau$ & re-scaled time near deterministic crossing time $t^*$ & $t=t^*+\frac{M^{\nhalf}}{-\mu(\ell,t^*)}\tau$ \\ \hline
\end{tabularx}
\caption{\emph{Summary of the different coordinate systems used to describe the population in the model of Sec.~\ref{ssec:model}.} Original coordinates (A) appear in the master equation~\eqref{eq:CME}, while coordinates (B) and (C) are used in the Kramers--Moyal expansion and linear-noise approximation, respectively [see Eqs.~\eqref{eq:SDE} and \eqref{eq:LNA}]. Coordinates (D) are used for our analysis of the dynamics in the narrow, boundary-crossing region. The subscript $t$ (or $\tau$) is used to denote random processes.}
\label{table:coordinates}
\end{table*}

In order to focus on the dynamics in a time window near $t^*$, it is useful to introduce re-scaled coordinates
\begin{subequations}\begin{align}
t={}&t^*-\frac{M^{\nhalf}}{\mu(\ell,t^*)}\tau, \\
 \quad n_\tau={}&\ell+M^{\nhalf} \zeta_\tau.
\label{eq:new_coords}
\end{align}\end{subequations}
Considering values of $\tau$ and $\zeta$ of order $M^0$ allows us to magnify the region around $t^*$ where boundary crossings are likely. In these coordinates, the crossing of the deterministic trajectory occurs at $\tau=0$, and the position of the threshold is at $\zeta=0$. We note that $\mu(\ell,t^*)<0$ so that positive values of the re-scaled time ($\tau>0$) correspond to $t>t^*$. A summary of the coordinates used in our analysis is given in Table~\ref{table:coordinates}.

Substituting the new coordinates into Eq.~\eqref{eq:FPE}, and writing $\widetilde \Pi(\zeta,\tau)$ for the probability density in these coordinates, we find
\BE
\frac{\partial}{\partial \tau}\widetilde\Pi(\zeta,\tau)&=&
\frac{1}{\mu(\ell,t^*)}\frac{\partial}{\partial \zeta}\left[\mu(\ell+M^{\nhalf} \zeta,t)\widetilde \Pi(\zeta,\tau)\right]\nonumber \\
&&+\frac{1}{\mu(\ell,t^*)}\frac{1}{2M^{\half}}\frac{\partial^2}{\partial \zeta^2} \left[\sigma^2(\ell+M^{\nhalf} \zeta,t)\widetilde \Pi(\zeta,\tau)\right].
\EE
Expanding in powers of $M^{\nhalf}$ we find to lowest order $\frac{\partial}{\partial \tau}\Pi(\zeta,\tau)=\frac{\partial}{\partial \zeta}\Pi(\zeta,\tau)$, i.e., near the threshold the dynamics of the system can be approximated by
\be
\zeta(\tau)=\zeta_0-\tau,
\label{eq:linear_path}
\ee
where $\zeta_0$ is the location of the path at time $\tau=0$ (i.e., at $t=t^*$). Fig.~\ref{fig:process}\,(b) shows a number of different stochastic trajectories in this region. Broadly, they travel along approximately parallel straight paths of gradient minus one (in the coordinate system of $\tau$ and $\zeta$).

We now use this result to approximate the distribution of crossing times. To do this we estimate when a particular trajectory located at $\xi_0$ at time $t^*$ crosses (or did cross) the threshold. We write $\tau_\times(\xi_0)$ for this crossing time in the re-scaled coordinates. Using Eq.~\eqref{eq:linear_path} we find
\begin{align}
\tau_\times(\zeta_0) = \zeta_0 \text{ .}
\label{eq:mapping}
\end{align}
We show this schematically in Fig.~\ref{fig:process}(c). We now combine this with the Gaussian distribution for $\xi_0$ obtained from the LNA, also shown in Fig.~\ref{fig:process}(c). Equation~\eqref{eq:dist}, evaluated at $t=t^*$, can be written as
\begin{align}
\Pi(\zeta_0)=&\frac{1}{\sqrt{2\pi\Sigma^2(t^*)}}\exp\left(-\frac{\zeta_0^2}{2\Sigma^2(t^*)}\right),
\end{align}
and we use this together with Eq.~(\ref{eq:mapping}) to approximate the distribution of first-passage times $t_\times$ as
\begin{align}
p(t_\times)=&\sqrt{\frac{M \mu^2(\ell,t^*)}{2\pi\Sigma^2(t^*)}} \exp\left(-\frac{M \mu^2(\ell,t^*)}{2\Sigma^2(t^*)} \left(t_\times-t^*\right)^2 \right).
\label{eq:FPT}
\end{align}
Using the definition of NTCP as outlined above we find 
\be\label{eq:erf}
\NTCP(t)=\frac{1}{2}\left[1+\erf\left(\frac{(t-t^*)\sqrt{M}\mu(\ell,t^*)}{\sqrt{2}\Sigma(t^*)}\right)\right],
\ee
where $\erf$ is the error function.

\begin{figure*}[t]
\begin{center}
\includegraphics[width=0.85\textwidth]{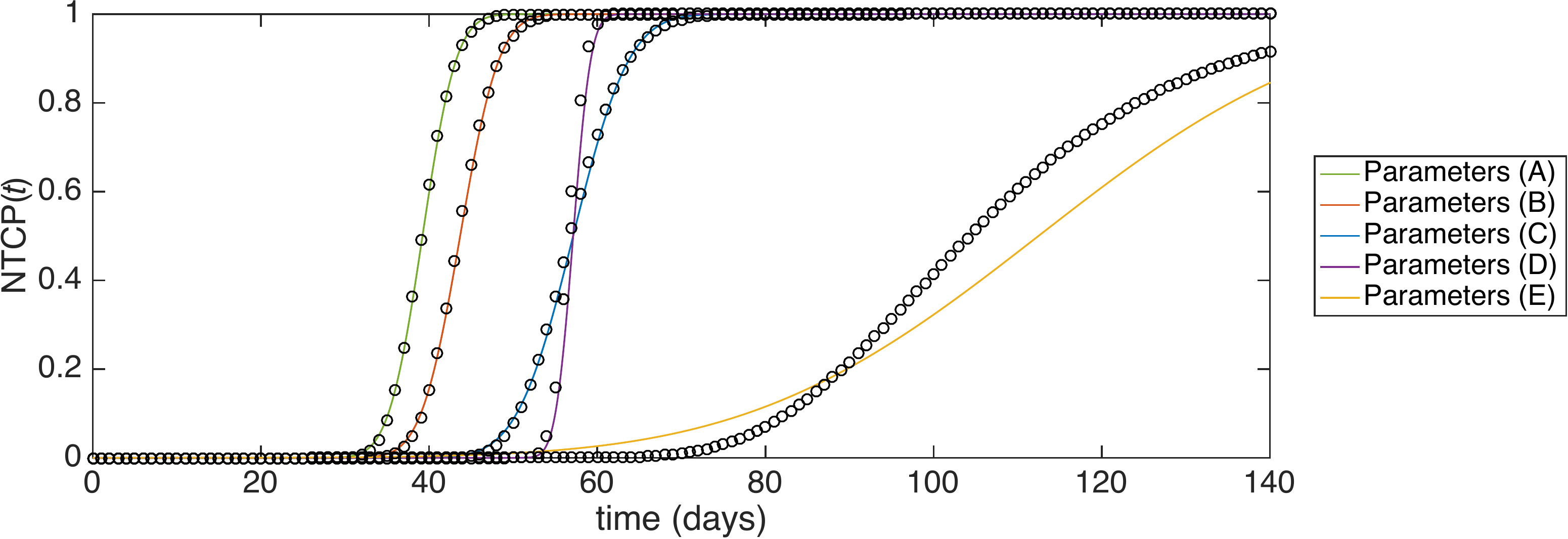}
\end{center}
\caption{\emph{NTCP as a function of time for the logistic model of healthy tissue in Sec. \ref{ssec:model}.} Black circles are obtained from numerical integration of the master equation of the original model [Eq.~\eqref{eq:CME}]. Coloured solid lines show the approximation of Eqs.~\eqref{eq:mean_fpt} and~\eqref{eq:var_fpt}.
Model parameters are given in Table~\ref{table:parameters_simple}.
}
\label{fig:NTCP_simple}
\end{figure*}

\begin{table*}[t!]
\small
\centering
\begin{tabular}{|l l l l l l l|} 
 \hline
 Parameter~& Definition & Value &&&& \\ [0.5ex]
 & & (A) & (B) & (C) & (D) & (E)\\
 \hline
 $b_0$ & mitosis rate (day${}^{-1}$) & 0.019~ & 0.019~ & 0.019~ & 0.019~ & 0.038~ \\
 $d$ & natural death rate (day${}^{-1}$) & 0.002 & 0.002 & 0.002 & 0.002 & 0.004\\ 
 $h_0$ & irradiated death rate (day${}^{-1}$) & $0.035$ & $0.032$ & $0.026$ & $0.026$ & $0.026$ \\
 $M$ & typical population size (see text) & $500$ & $500$ & $500$ & $5000$ &$500$ \\
 $\ell$ & threshold for onset of NTC~& $1/3$ & $1/3$ & $1/3$ & $1/3$ & $1/3$ \\
 \hline
\end{tabular}
\caption{\emph{Five sets of parameters used in Fig.~\ref{fig:NTCP_simple} for the logistic model of healthy tissue.} These parameter sets are the same as those considered in Ref.~\cite{stocks2016stochastic}, but we have defined separate mitosis and natural death rates to be able to analyse stochastic effects in finite populations (see text). The ratio of mitosis and natural death was chosen as $10:1$, consistent for example with Ref.~\cite{hanin2013mechanistic}.}
\label{table:parameters_simple}
\end{table*}

\subsubsection{Closed-form approximation of NTCP for model with logistic growth and constant radiation}
\label{sec:closed_form}
We now test this approximation scheme on the logistic growth model defined in Eq.~(\ref{eq:reactions_simple}). We focus on a particularly simple case where there is no radiation prior to a certain time, and a constant rate of death due to radiation thereafter. We choose time $t=0$ as the point at which radiation sets in, so that the hazard function $h(t)$ is the step function
\begin{equation}
h(t)=\left\{\begin{array}{c}
0 \quad\text{for $t<0$},\\
h_0 \quad\text{for $t\geq 0$}.
\end{array}\right.
\end{equation}
We primarily consider radiation of this type as a simple initial example, following the study of NTCP in Ref.~\cite{stocks2016stochastic}. More complicated radiation protocols will be discussed below.

We assume that the dynamics of the population start long before $t=0$, so that the stationary state of the master equation~(\ref{eq:CME}) [with $h(t)=0$] is reached by $t=0$. The mean and variance of this distribution are given by the fixed points of Eqs.~(\ref{eq:LNA_top}) and (\ref{eq:variance}), using $\mu$ and $\sigma^2$ for the logistic model and setting $h(t)=0$. We have
\begin{subequations}\begin{align}
\phi(t=0) ={}& 1, \label{eq:phi0}\\
\Sigma(t=0) ={}& \frac{d}{b_0-d} \label{eq:sig0}.
\end{align}\end{subequations}
At times $t\geq 0$, Eqs.~\eqref{eq:LNA_top} and~\eqref{eq:variance} are given by 
\begin{subequations}\begin{align}
\frac{\d \phi}{\d t} ={}& \phi b_0\left(1-\frac{\phi}{k}\right)-\phi\left[d+h_0\right], \label{eq:LNA_logistic_top}\\
\frac{\d \Sigma^2}{\d t} ={}& 2 \left\{ b_0\left(1-\frac{2\phi}{k}\right)-\left[d+h_0\right] \right\} \Sigma^2 + \phi b_0\left(1-\frac{\phi}{k}\right)+\phi\left[d+h_0\right].\label{eq:LNA_logistic_bottom}
\end{align}\end{subequations}
Eq.~\eqref{eq:LNA_logistic_top} can be solved in closed form subject to the initial condition $\phi(0)=1$. From the resulting deterministic trajectory $\phi(t)$ one then finds the passage time $t^*$ of the deterministic trajectory as 
\begin{equation}
t^*=\frac{1}{b_0-d-h_0}\log \left(\frac{h_0 \ell}{b_0\ell - d\ell -b_0 +d + h_0}\right),
\label{eq:mean_fpt}
\end{equation}
assuming the fixed point of the deterministic trajectory is below the boundary $\ell$.
Next we turn to Eq.~\eqref{eq:LNA_logistic_bottom} in order to find $\Sigma^2(t^*)$. For constant radiation the path $\phi(t)$ is monotonically decreasing in time. This allows us to trade the time derivative in Eq.~(\ref{eq:LNA_logistic_bottom}) for a derivative with respect to $\phi$, resulting in a linear ODE for $\Sigma^2$ as a function of $\phi$.
For our specific example this ODE can be solved in closed form, and we find the variance of first-passage times as
\begin{align}\begin{split}
\frac{\Sigma^2\left(t^*\right)}{M\mu^2(\ell,t^*)}=&\frac{
5b
+\frac{2(b_0-d)d}{h_0}
+\frac{(b_0-2d)h_0}{b_0-d}
-\frac{b_0+d+h_0}{\ell}
+ \frac{(b_0-d)(b_0-d-h_0)(d+h_0)}{[d+h_0+b_0(\ell-1)-d\ell]^2}
- \frac{(b_0-d)[b_0+3(d+h_0)]}{d+h_0+b_0(\ell-1)-d \ell} }{M(b_0-d-h_0)^3}
\\&+
\frac{2(b_0-d)(b_0+2d+2h_0) \log\left(\frac{h_0 \ell}{b_0\ell-d\ell-b_0+d+h_0}\right)}{M(b_0-d-h_0)^4}.
\label{eq:var_fpt}
\end{split}\end{align}
 This can then be used in Eq.~(\ref{eq:erf}) to obtain $\NTCP(t)$.

In Fig.~\ref{fig:NTCP_simple} we show the resulting NTCP as a function of time for several sets of model parameters; these parameter sets are summarised in Table~\ref{table:parameters_simple}, and were previously motivated and used in Ref.~\cite{stocks2016stochastic} to consider normal tissue complications arising from the treatment of prostate cancer. In order to test the accuracy of our approximation, we have also obtained NTCP$(t)$ for the original model by numeral integration of the master equation Eq.~\eqref{eq:CME}; these values are shown as black circles in Fig.~\ref{fig:NTCP_simple}. These results are compared with the analytical approximations in Eqs.~\eqref{eq:erf} and~\eqref{eq:var_fpt}, and for most of the parameter sets tested we find good agreement. The approximation works noticeably less well for parameter set (E) than for the other four sets. In this case, the speed with which the deterministic path crosses the boundary is lower than for the other parameter sets. This leads to a longer time window around $t^*$ within which crossings are likely, and thus a larger amount of error in our approximation.

\section{Extended model of normal and doomed cells}
\label{sec:2d}
\subsection{Model definitions}\label{ssec:model2d}

Hanin and Zaider~\cite{hanin2013mechanistic} proposed a model which adds complexity by including radiation-damaged cells.
In this model, damaged cells continue to occupy the limited volume available to the population. Damaged cells also carry out their functions, but fail to proliferate.
The presence of such cells has been offered an explanation for the observation that, after irradiation, an initial lag period occurs before re-population \cite{hanin2013mechanistic,hall2006radiobiology}.
Similar models have been proposed for tumour cells for a more realistic calculation of TCP, where the population is divided into radiation-damaged and unaffected tumour cells \cite{ponce2017stochastic}.

As before there are `normal cells' $\mathcal{N}$ which carry out the functions of the organ; these cells have the ability to proliferate.
However, once damaged by radiation, a cell does not vanish immediately; rather, it becomes a `doomed cell' $\mathcal{X}$ \cite{hanin2013mechanistic}.
Doomed cells continue to contribute to the normal functions of the organ, however they are unable to proliferate.
Thus, although they may temporarily aid the function of the organ, they ultimately die without reproduction.
Doomed cells also consume resources and so are in direct competition with the normal cells. As a result of this, the per capita mitosis (birth) rate of normal cells decreases as the total size of the population of both types increases. The dynamics of the model can be summarised as follows:
\begin{align}
\begin{aligned}[c]
{\mathcal N} \xrightarrow{\mathmakebox[20mm]{ b_0 \left(1 - \tfrac{N+X}{k M} \right) }}{}& {\mathcal N}+ {\mathcal N} \quad&& \text{(mitosis of normal cells)}, \\
{\mathcal N} \xrightarrow{\mathmakebox[20mm]{ h (t) }}{}&\mathcal{X} &&\text{(radiation damage)}, \\
{\mathcal N} \xrightarrow{\mathmakebox[20mm]{ d_1 }}{}& \emptyset &&\text{(death of normal cell)}, \\
{\mathcal X} \xrightarrow{\mathmakebox[20mm]{ d_2 }}{}& \emptyset &&\text{(death of doomed cell)}.
\end{aligned}
\label{eq:reactions_2d}\end{align}
We write $N$ and $X$ for the numbers of normal and doomed cells, respectively. As before, the constant $k\equiv\left(1-d_1/b_0\right)^{-1}$ is chosen so that---in the absence of radiation---the stationary average size of the population of normal cells is $M$. An NTC is assumed to arise when the total number of functional cells, $N+X$, falls below a threshold $L$.

Writing $s=(N+X)/M$ for the (re-scaled) total number of functional cells in the population, and $x=X/M$ for the (re-scaled) number of doomed cells, one has the following rate equations in the deterministic limit,

\begin{subequations}\begin{align}
\frac{\d s}{\d t}=& b_0\left(1-\frac{s}{k}\right)(s-x) - d_1 (s-x) - d_2 x, \\
\frac{\d x}{\d t}=&h(t) (s-x) - d_2 x.\label{eq:det2d}
\end{align}\end{subequations}

In this example, we consider brachytherapy where there is a time-varying dose of radiation acting on the population of normal cells, resulting from the decay of a radioactive implant. The effect of this type of radiation on the population of normal cells is obtained using the linear-quadratic (LQ) formalism, which is well established in the modelling of brachytherapy \cite{brenner1997use,brenner2008linear,fowler201021}.
This formalism accounts for the degradation of the radioactive implant, both linear and quadratic tissue responses to radiation, and DNA repair. This leads to a time-dependent radiation hazard rate for the conversion of normal cells into doomed cells:
\begin{equation}
h(t)=\alpha R_0 e^{-\lambda t} + \frac{2 \beta R_0^2 e^{-\lambda t} }{\gamma - \lambda} \left( e^{-\lambda t} - e^{- \gamma t}\right),
\label{eq:h_brachymain}
\end{equation}
where $\alpha, \beta, \gamma, \lambda$ and $R_0$ are model parameters; $R_0$ in particular denotes the initial dose rate. Further details are given in \ref{app:LQ}. We consider a specific set of realistic parameters, proposed by Hanin and Zaider \cite{hanin2013mechanistic} and summarised in Table~\ref{table:parameters_2d}. These parameters were chosen to model the treatment of prostate cancer, where the normal-tissue complication refers to grade $2$, or larger, toxicity (`GU2+') of the genitourinary tract.
\begin{figure*}[t]
\centering
\includegraphics[width=0.85\textwidth]{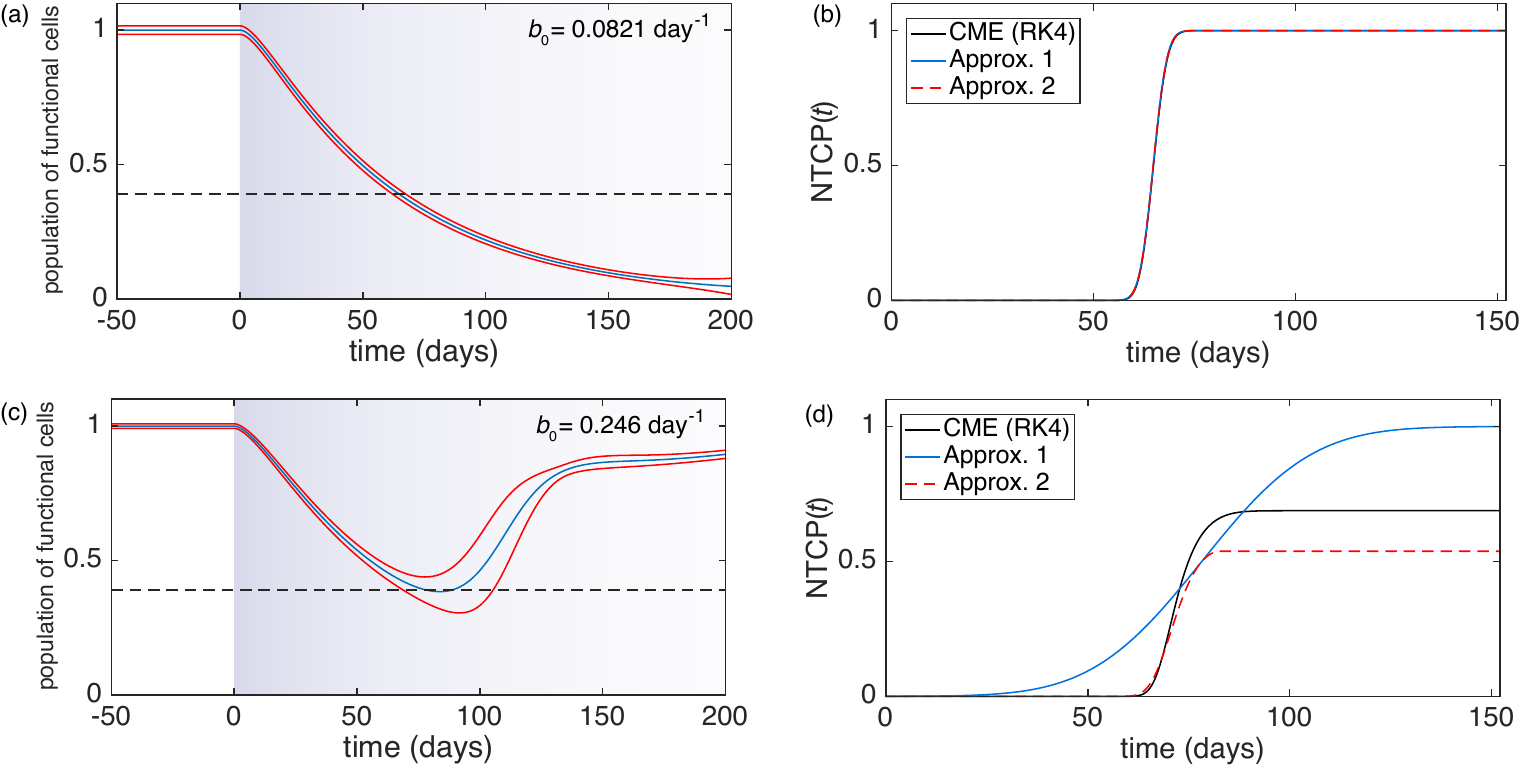}
\caption{\emph{Behaviour of the model with normal and damages cells defined in Sec.~\ref{ssec:model2d}}. Panels (a) and (c): Population density for number of functional cells as a function of time for two different parameter sets (see Table~\ref{table:parameters_2d}). The central blue line shows the deterministic trajectory [Eq.~\eqref{eq:det2d}], red lines indicate a band of one standard deviation as predicted by the linear-noise approximation. The shading of the background indicates the rate of radiation damage $h(t)$.
Panels (b) and (d): NTCP as a function of time. We compare the results of our two approximations with the outcome of numerical integration of the (chemical) master equation (CME) using a Runge--Kutta scheme (RK4).
 }
\label{fig:2d}
\end{figure*}

\begin{table*}
\small
\centering
 \begin{tabular}{|l l l l|} 
 \hline
 Parameter~ & Definition & Fig.~\ref{fig:2d}\,(a, b)~ & Fig.~\ref{fig:2d}\,(c, d)\\
 \hline
 $b_0$ & mitosis rate (day${}^{-1}$) & 0.0821~ & 0.246~ \\
 $d_1$ & normal cell death rate (day${}^{-1}$) & 0.0164 & 0.0164 \\ 
 $d_2$ & irradiated cell death rate (day${}^{-1}$) & 0.0164 & 0.0164 \\ 
 $M$ & population size & $1000$ & $1000$ \\
 $\ell=\tfrac{L}{M}$ & critical fraction of population~& $0.39$ & $0.39$ \\ 
 $ \alpha $ & LQ model parameter ($\text{G}\,\text{y}^{-1}$)& 0.109 & 0.109 \\
 $ \beta $ & LQ model parameter ($\text{G}\,\text{y}^{-2}$)& 0.0364 & 0.0364\\
 $ \gamma $ & rate of DNA repair ($\text{month}^{-1}$) & 720 & 720 \\
 $ R_0 $ & initial dose rate of implant ($\text{G}\,\text{day}^{-1}$) & 1.68 & 1.68 \\
 $ \lambda $ & decay rate ($\text{day}^{-1}$) & 0.0117 & 0.0117 \\
 \hline
\end{tabular}
\caption{\emph{Parameters used in Fig.~\ref{fig:2d}.} Similar parameters were previously proposed in Ref.~\cite{hanin2013mechanistic}. We have explicitly included normal-cell birth and death and made the assumption that $d_1=d_2$.}
\label{table:parameters_2d}
\end{table*}
 
 \subsection{Alternative approximation for NTCP}\label{sec:approx2}
Results for this model are presented in Fig.~\ref{fig:2d}. We first focus on the deterministic dynamics, indicated by the blue lines in panels (a) and (c). In panel (a) the mitosis rate $b_0$ is sufficiently low for deterministic trajectory to fall below the threshold $\ell$ for the onset of NTCs. The approximation for NTCP developed in Sec.~\ref{ssec:approx1} can be applied, as discussed in more detail in Sec.~\ref{ssec:certain}. 

The second parameter set in Table \ref{table:parameters_2d} describes a case with a higher mitosis rate $b_0$. As shown in Fig.~\ref{fig:2d} (c), the solution of the deterministic rate equations then only briefly falls below the threshold $\ell$. The number of functional cells then increases again to values above $\ell$. In the stochastic system we expect only a fraction of trajectories to cross the threshold; some realisations may never fall below $\ell$, and hence $\NTCP(t)$ can be expected to take a long-time limit below one. This cannot be captured by the approximation method in Sec.~\ref{ssec:approx1}.

With this in mind, we propose the following improved method of estimating NTCP.
Within the LNA, at each moment in time~$t$ the distribution of the population of interest (in this case $s_t$) is approximately normal with a mean $\phi(t)$ and variance $\Sigma^2(t)$ given by Eqs.~\eqref{eq:LNA_top} and~\eqref{eq:variance}, respectively.
The amount of probability below the threshold $\ell$ at a given time is then obtained as\footnote{We note that the quantity $Q(t)$ in Eq.~\eqref{eq:q} corresponds to NTCP as defined in Ref.~\cite{stocks2016stochastic}.}
\begin{equation}\label{eq:q}
Q(t)=\frac{1}{2}\left[1+\erf\left(\frac{\sqrt{M}[\ell-\phi(t)]}{\sqrt{2} \Sigma(t)}\right)\right].
\end{equation}

We now estimate $\NTCP(t)$ as the maximum amount of probability below the threshold at any earlier time $t' \leq t$, i.e.,
\begin{equation}\label{eq:maxq}
\text{NTCP}(t)=\max_{t' \leq t } \,Q(t') .
\end{equation}
Further steps of the mathematical evaluation are presented in \ref{sec:2d_calculation}. 

We briefly comment on the limitations of this approximation, before we discuss the results for the model of normal and doomed cells.
Equation~(\ref{eq:maxq}) provides a lower bound for NTCP of the process described by the LNA.
This can be seen as follows.
At a given time $t$, let the maximum in Eq.~(\ref{eq:maxq}) have occurred at a time $t_m\leq t$;
the estimate for $\NTCP(t)$ is then $Q(t_m)$.
Consider now a trajectory with a total population density above the boundary at time $t_m$, $s_{t_m}>\ell$.
Such a trajectory does not contribute to $\NTCP(t)$ within our approximation, even though it may have well have attained population sizes below threshold before $t_{\rm m}$, or go below threshold between $t_m$ and $t$. The above approximation therefore underestimates NTCP.
We note that the SDE obtained in the LNA is itself an approximation, so the above calculation is not necessarily a lower bound to the NTCP of the discrete population dynamics from which we started.

 \begin{figure*}
\begin{center}
\includegraphics[width=0.60\textwidth]{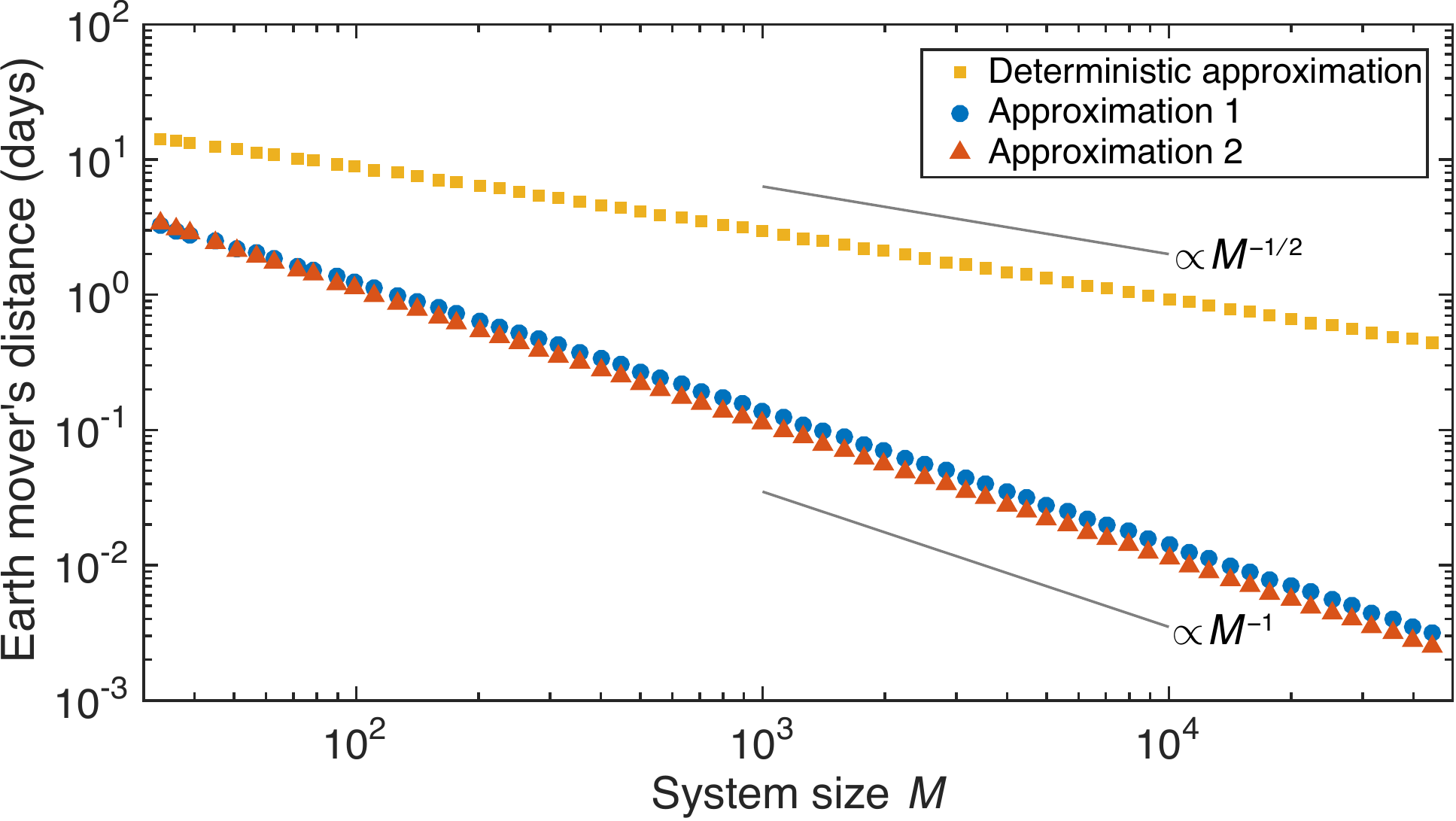}
\end{center}
\caption{\emph{Measure of error for the predictions of NTCP for the model in Sec.~\ref{ssec:model}}. We use the Earth Mover's Distance (EMD) \cite{gibbs2002choosing} as a measure of distance between two probability distributions. Each set of symbols shows the EMD of the distribution of first-passage times obtained from the different approximations relative to the distribution obtained for the original model obtained by numerical integration of the master equation~(\ref{eq:CME}). We compare three approximations: the deterministic approximation from Ref.~\cite{stocks2016stochastic} (i.e., the distribution of first-passage times is a delta-peak at the deterministic crossing time $t^*$ ), and Approximations $1$ and $2$ as described in the text. Results are shown as a function of the population-size parameter $M$. The data indicates that the EMD of Approximations $1$ and $2$ from the original model scales as $M^{-1}$ with the typical size of the population; similar scaling is also observed using the Kullback--Leibler divergence (not shown). For the deterministic approximation the EMD decays much more slowly with the system-size parameter ($\propto M^{-1/2}$).}
\label{fig:divergence}
\end{figure*}
Despite these limitations, the method provides useful estimates for NTCP. For example, $\NTCP(t)$ obtained from Eqs.~(\ref{eq:q}) and (\ref{eq:maxq}) for the model in Sec.~\ref{ssec:model} does not significantly differ from the predictions of the method discussed in Sec.~\ref{ssec:approx1}. To keep the language compact we will refer to the procedure in Sec.~\ref{ssec:approx1} as Approximation $1$ from now on, and to that in Eqs.~(\ref{eq:q}) and (\ref{eq:maxq}) as Approximation $2$. A quantitative comparison of the distributions of first-passage time from the two approximations for the model in Sec.~\ref{ssec:model} is shown in Fig.~\ref{fig:divergence}. The data indicates that Approximation $2$ provides an improvement relative to Approximation $1$. Both methods do considerably better than the deterministic approximation in Ref.~\cite{stocks2016stochastic}.

To compare the three approximations we have use the Earth-Movers distance (EMD), also known as the Wasserstein metric \cite{gibbs2002choosing}. Intuitively, it is a measure of the amount of `effort' needed to turn one distribution into the other; it is
the amount of probability that needs to be moved weighted by the distance it has to be moved. We choose this rather than, say, the Kullback--Leibler divergence \cite{kullback1951information} or total variation distance since the distribution of first-passage times from the deterministic approach is a Dirac delta-distribution \cite{stocks2016stochastic} which results in infinite Kullback--Leibler divergence. The EMD gives a more useful measure of error.

 \subsection{NTCP for model of normal and doomed cells}
 For the model with normal and doomed cells Approximation $2$ can provide a significantly improved prediction of NTCP compared to Approximation $1$, as we will discuss in this section. In this context it is useful to distinguish the cases in which normal tissue complication occurs with certainty at long times and those in which long-time NTCP stays below one.

\subsubsection{Certain normal tissue complication at long times}\label{ssec:certain}
For the first set of parameters in Table~\ref{table:parameters_2d} normal-tissue complication occurs with probability one at long times. We show results in panel (a) of Fig.~\ref{fig:2d}. The source of radiation is implanted at time zero, assuming that the population of normal cells is at its stationary state at this time. The population of functional cells then decreases monotonously, and the number of functional cells crosses the threshold for the onset of NTC. Panel (b) shows the estimates for $\NTCP$ as a function of time for Approximation $1$ and Approximation $2$. Their predictions are largely indistinguishable, and they both agree well with results for the original model found by numerical integration of the master equation.

We note that for this choice of parameter values, carrying out the numerical integration of the master equation takes approximately $10^5$ times longer than to evaluate each of the two approximations. This is because the master equation consists of a set of $M^2$ coupled ODEs, whereas evaluation of each of the approximations only involves integrating forward five ODEs (for the means of the two degrees of freedom, their variances and the covariance). Thus, the approximation methods offer a significant increase in efficiency for large populations, at moderate reduction of accuracy.

\subsubsection{Uncertain onset of normal tissue complication}
In panels (c) and (d) of Fig.~\ref{fig:2d} we show the same quantities, but for a different choice of birth rate (see Table \ref{table:parameters_2d}). The deterministic path barely crosses the boundary $\ell$, and for this choice of parameters only a fraction of trajectories of the stochastic model will lead to an onset of NTC. In this case, the predictions of the two approximations are widely different. Approximation $1$ assumes a Gaussian distribution of first-passage times and deviates significantly from the NTCP seen in the original model. Most notably, this approximation predicts that all trajectories eventually cross the boundary so that $\NTCP(t)\rightarrow 1$ at large times. Although this is not the case for typical population size used in this example ($M=1000$), we remark that for $M\to\infty$ NTC becomes certain at long times in the original model for the present parameter set. 

As seen in Fig.~\ref{fig:2d} (d) Approximation $2$ outperforms Approximation $1$. This is because, in the narrow region where boundary-crossings are likely, there is a significant change in the drift for the total population size; the sign of the drift changes from negative to positive. Approximation $2$ takes this into account, whereas Approximation $1$ is based on constant drift within the region near the boundary $\ell$.
Unlike Approximation 1, Approximation 2 does not (wrongly) predict that all trajectories eventually cross the boundary. Instead $\NTCP(t)$ remains below unity at $t\to\infty$ within Approximation $2$.

\section{Complication-free tumour control}\label{sec:CFC}
\subsection{Motivation}
The objective of radiation therapy is to successfully eliminate cancerous cells while avoiding further complications from damaging normal tissue cells.
In the preceding sections, we outlined analytical approximations for the efficient calculation of NTCPs. Tumour control probabilities---the probability of eliminating all cancer cells---from a stochastic birth-death model have been previously considered by Zaider and Minerbo~\cite{zaider2000tumour}; the authors derive a general equation for the probability of the elimination of all tumour cells.
In this section, we combine these two results for NTCP and TCP respectively to investigate how, in principle, mathematical models can be used to optimise the application of radiation therapy to achieve complication-free tumour control.
We begin by motivating an extension to the model described in Sec.~\ref{sec:model} to include the growth of cancerous cells.
For completeness, we then proceed by briefly reviewing Zaider and Minerbo's result describing TCP.
\subsection{Model definitions}\label{ssec:cfcmodel}
We consider a model which contains both normal cells $\mathcal{N}$ and cancerous cells $\mathcal{C}$. The two populations are assumed to be spatially separated from each other. The normal cells are as described in Sec.~\ref{sec:model}: they undergo mitosis with a rate which depends on the number of normal cells, leading to logistic growth. They are also subject to a natural death with rate $d_1$, and to death from a source of radiation with hazard function $h_1(t)$. We label the rates pertaining to normal cells with the subscript $1$, and similarly subscript $2$ for cancerous cells. Cancerous cells, on the other hand, undergo mitosis with a constant rate $b_2$ \cite{zaider2000tumour}; numerical evidence suggests that the resulting exponential growth characterise tumours of small sizes well \cite{mcaneney2007investigation}. Cancer cells are also subject to a natural death with a rate $d_2$ and to death from a source of radiation with hazard function $h_2(t)$.
The model can be summarised by the following reactions:
\begin{align}
\begin{aligned}[c]
\mathcal{N} \xrightarrow{\mathmakebox[15mm]{ b_1 \left(1 - \tfrac{N}{k M} \right) }}{}& \mathcal{N}+\mathcal{N},\quad & \mathcal{C} \xrightarrow{\mathmakebox[15mm]{ b_2 }}{}& \mathcal{C}+\mathcal{C},\quad && \text{(mitosis)},\\
\mathcal{N} \xrightarrow{\mathmakebox[15mm]{ d_1 }}{}& \emptyset,\quad & \mathcal{C} \xrightarrow{\mathmakebox[15mm]{ d_2 }}{}& \emptyset, && \text{(natural death)},\\
\mathcal{N} \xrightarrow{\mathmakebox[15mm]{ h_1(t) }}{}& \emptyset,\quad & \mathcal{C} \xrightarrow{\mathmakebox[15mm]{ h_2(t) }}{}& \emptyset, && \text{(irradiated death)}.
\end{aligned}
\label{eq:reactions_both}\end{align}
Although both cells are subject to the same source of radiation, the hazard functions $h_1(t)$ and $h_2(t)$ for the two cell types can differ.
This is because each cell type differs in its susceptibility to radiation and in their ability to repair damaged DNA.
We again consider the case of brachytherapy, as in Sec.~\ref{sec:2d}. The hazard function is as in Eq.~\eqref{eq:h_brachymain}, where the parameters $\alpha_{1,2}$, $\beta_{1,2},$ and $\gamma_{1,2}$ depend on the cell type.
We also assume that, due to the presumed spatial separation of normal tissue and cancerous cells, the treatment can be targeted such that each cell type absorbs a different fraction of the total dose rate.
This is incorporated into the hazard function by replacing the initial dose rate $R_0$ with an effective dose rate $\theta_{1,2}R_0$.
The parameters describing the initial dose rate $R_0$ and the decay rate $\lambda$ are characteristics of the radioactive implant and are thus common to the hazard function of both cell types.
As before, we initialise the population of normal cells in its stationary state. We let there be initially $C_0$ cancer cells. 

\begin{figure*}
\centering
\includegraphics[width=0.8\textwidth]{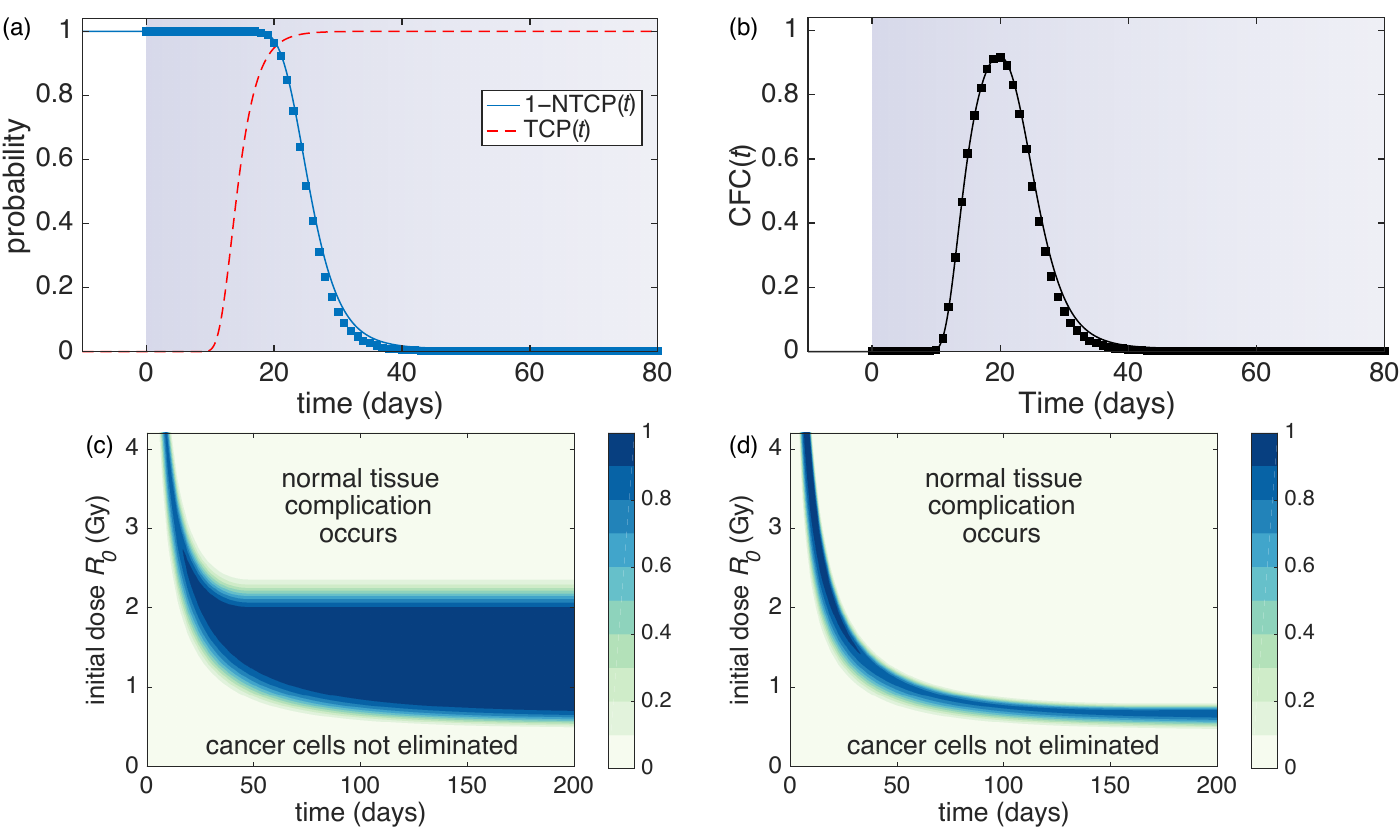}
\caption{\emph{TCP, NTCP and probability of CFC for the model in Sec.~\ref{ssec:cfcmodel}.} Panel (a): Probability that normal tissue complication has not yet occurred, $1-{\rm{}NTCP}(t)$, as predicted by Approximation $2$ (blue line) and from numerical integration of the master equation (blue squares). Probability that the tumour is successfully eliminated ${\rm TCP}(t)$ (dashed red line). TCP is calculated as in Ref.~\cite{zaider2000tumour}. The shading of the background indicates the hazard function $h(t)$. Initial dose $R_0=2.5\text{G\,y}$.
Panel (b): Resulting probability of complication-free tumour control ${\rm CFC}(t)$. Black line is using Approximation $2$ for NTCP and TCP as in Ref.~\cite{zaider2000tumour}; results from direct numerical integration of the master equation are shown as black squares.
Panels (c) and (d): CFC$(t)$ for different values of the initial dose $R_0$, and for two different sets of model parameters (see Table~\ref{table:CFC}).
}
\label{fig:NTCPvsTCP}
\end{figure*}

\subsection{Tumour control probability, normal-tissue complication probability, and probability of complication-free control}
We now consider the probability as a function of time of eliminating all cancer cells--- ${\rm TCP}(t)$.
Similarly to the calculation of ${\rm NTCP}(t)$, this is mathematically a first-passage time problem. Zaider and Minerbo \cite{zaider2000tumour} developed an analytical description for TCP for the linear dynamics of cancerous cells described above.
This was achieved using a generating-function. This approach is feasible due to two features of the problem: (i) the model is linear (i.e., cells do not interact with each other), and (ii) the boundary of interest for TCP is at zero (i.e., extinction of tumour cells). The result for TCP$(t)$ is \cite{zaider2000tumour}
\begin{align}
{\rm TCP}(t)=\left[ 1 - \frac{C(t)/C_0}{1+b_2 \int_0^t \d t' \frac{C(t)}{C(t')}}\right]^{C_0},
\label{eq:TCP}\end{align}
where $C(t)$ is the deterministic path for number of cancerous cells, given by
\begin{align}
\frac{\d C}{\d t} = \left[b_2-d_2-h_2(t)\right]C(t).
\label{eq:mean_field_cancer}\end{align}
Even though the expression involves the deterministic trajectory $C(t)$ we stress again that this result is exact for arbitrary population sizes, and does not imply any approximations. While Eq.~\eqref{eq:mean_field_cancer} cannot be solved analytically in most cases, the equation can be integrated numerically for an efficient calculation of TCP$(t)$. The analysis reviewed here has also been extended to consider more complicated models, including the different stages of the cell cycle \cite{dawson2006derivation,maler2009cell,hillen2010cell}.

Complication-free tumour control (CFC) refers to the elimination of all cancer cells while maintaining enough normally functioning tissue for an organ to operate without complications \cite{mundt2005intensity}. The probability of CFC as a function of time is therefore given by \cite{mundt2005intensity}
\begin{equation}
{\rm CFC}(t) = {\rm TCP}(t)\left[1-{\rm NTCP}(t)\right].
\label{eq:CFC}
\end{equation}
We remark that Eq.~\eqref{eq:CFC} implies an equal weighting of the importance of tumour control and NTCs.
In the most extreme cases, for example where NTCs relate to organ failure this is justified.
In other cases, for example when NTC refers to increased urinal frequency, a complication may be preferable to a potentially life-threatening tumour.
In such cases, Eq.~\eqref{eq:CFC} can be modified by appropriately weighting the two probabilities to maximise a `quality of life' measure in accordance with clinical experience \cite{kallman1992algorithm}.

Fig.~\ref{fig:NTCPvsTCP}\,(a) shows the probabilities $1-{\rm NTCP}(t)$ and ${\rm TCP}(t)$ for the model defined in Sec.~\ref{ssec:cfcmodel}, and for a specific choice of parameters (see Table~\ref{table:CFC}).
These quantities are obtained by Approximation 2 for NTCP, and Eq.~\eqref{eq:TCP} for TCP.
Similarly, Fig.~\ref{fig:NTCPvsTCP}\,(b) shows ${\rm CFC}(t)$ and compares the results from our approximation to those of numerical integration of the master equation.
For this choice of parameters we find a non-trivial time ($\sim 20~{\rm days}$) which maximises the probability of CFC. In the case of a temporary brachytherapy implant, this would indicate the optimum moment for removal. 

The analysis provided here allows us to investigate the optimum application of brachytherapy to maximise the likelihood of CFC.
We consider a fixed set of parameters describing the cellular birth rates, death rates, susceptibilities and repair rates, shown in Table~\ref{table:CFC}.
We consider a temporary implant of a certain radioisotope, ${}^{125}\rm I$, which has a decay rate of $\lambda=0.0117~\text{day}^{-1}$.
In order to achieve CFC, we assume we are able to control the initial dose rate $R_0$ (i.e., the size of the radioactive seed) and the time at which the implant is removed.
\begin{table*}
\centering
\begin{tabular}{|l c c c c c c c c|} 
 \hline
 Case & $b$ & $d$ & $\alpha$ & $\beta$ & $\gamma$ & threshold &frac. of dose& pop. size\\
 & day${}^{-1}$ & day${}^{-1}$ & G\,y${}^{-1}$ & G\,y${}^{-2}$ & day${}^{-1}$ & for NTC $\ell$ & absorbed $\theta$ &\\
 \hline
 TCP Fig.~\ref{fig:NTCPvsTCP}\,(a--c) & 0.0165 & 0.0015 & 0.2 & 0.05 & 8.35 & ${}$ & 1.0& $C_0=10^3$\\
 NTCP Fig.~\ref{fig:NTCPvsTCP}\,(a--c) & 0.055 & 0.005 & 0.1 & 0.01 & 8.35 & 0.5 & 0.2& $M=10^3$\\ 
 \hline
 TCP Fig.~\ref{fig:NTCPvsTCP}\,(d) & 0.02 & 0.005 & 0.2 & 0.05 & 2.27 & & 1.0 & $C_0=10^3$\\
 NTCP Fig.~\ref{fig:NTCPvsTCP}\,(d) & 0.0067 & 0.0017 & 0.1 & 0.01 & 2.27 & 0.2 & 0.4& $M=10^3$\\ 
  \hline
\end{tabular}
\caption{Parameters used in Fig.~\ref{fig:NTCPvsTCP}, along with $\lambda=0.0117~\text{day}^{-1}$. The parameters in the upper two rows were previously used to model brachytherapy as a treatment for prostate cancer, where the normal tissue complication refers to rectal proctitis~\cite{stocks2016stochastic}. The parameters in the bottom row are hypothetical, used to show that a change in the optimum treatment strategy may result upon variation of parameters.}
\label{table:CFC}
\end{table*}

Fig.~\ref{fig:NTCPvsTCP}\,(c) shows the probability of CFC for different values of time and initial dose, again efficiently generated using Approximation 2 for NTCP and Eq.~\eqref{eq:TCP} for TCP.
With the exception of the population sizes, the parameters we choose here were previously used to model the treatment of prostate carcinoma~\cite{stocks2016stochastic} consistent with experimentally collected parameters~\cite{carlson2004comparison}. In this context NTC refers to acute radiation proctitis~\cite{kishan2015late}. For these parameters, the optimal strategy involves an initial dose of size $1.7~\text{G\,y}$ and removal at a time over $50~\text{days}$.
Using this initial dose, the probability of CFC$(t)$ does not decrease at large times, providing a large window for the removal of the implant or allowing the use of a permanent implant. This is not the case for all parameters; the optimum strategy may require the timely removal of the implant.
An example of this is shown in Fig.~\ref{fig:NTCPvsTCP}\,(d), which shows CFC$(t)$ for parameters where the cancer cells have a three-fold higher growth rate than normal cells. The probability of CFC is peaked when implanting a high dose of radiation for a short time.
For this case, we see the band where CFC is likely is narrow, indicating that such a treatment may be very sensitive to the time of removal of the implant.

\section{Conclusions}\label{sec:concl}
To summarise, we have derived approximations for the distribution of first-passage times through a boundary of a stochastic birth-death model. These approximations capture effects of fluctuations in the population discarded in previous approaches. The improvements rely on an expansion in the inverse typical size of the population. One can therefore expect the approach to be particularly useful for large, but finite populations. Intrinsic noise is then weak, but not always weak enough to be ignored altogether. It is worth noting that the methods we have developed do not require the birth-death model to be linear, for example we have considered logistic growth. Our analysis was presented in the context of normal tissue complication probabilities for radiotherapy treatment, however these mathematical results may also have wider applicability to other problems in which first-passage times of stochastic processes are of interest \cite{metzler2014}.

We note that NTCP takes the form of an error function in our approximation. This functional form has previously been reported in statistical models of NTCP, see for example Ref.~\cite{Lyman1985}. This indicates that NTCP can be different from zero or one for intermediate doses of radiation; NTC then occurs (or does not occur) as a random process. This is the case as well in our model; the source of stochasticity is the intrinsic noise in the population of functional cells, i.e., random birth and death events. It is not clear however what exactly the origin of uncertainty is in statistical models of NTCP. Intrinsic stochasticity within functional subunits, or resulting from small numbers of stem cells may be potential sources of randomness, but other factors are likely to contribute as well.

We have obtained approximations of NTCP for models of normal tissue with a single type of cell and for an extended model with two different cell types. Our results demonstrate that these approximations can lead to a significant increase in efficiency over simulation methods, at a moderate loss of accuracy. This is the case particularly when the underlying model becomes complex and has many different internal states. In the final part of the paper we showed how approximations of NTCP and TCP can be used to estimate the probability of complication-free tumour control. We have demonstrated how the analytical approximations can be used for the efficient identification of optimised parameters for treatment planning in brachytherapy. Our analysis is limited to stylised models, and we do not claim direct clinical applicability. However, we hope that the methods we have developed can be adapted to more realistic populations of cancerous cells and normal tissue.

\section*{Acknowledgements}
We acknowledge funding by the Engineering and Physical Sciences Research Council (EPSRC, UK) under grant numbersEP/K037145/1, EP/N033701/1 and in form of a studentship to PGH. We thank Thomas House and Oliver Jensen for discussions.

\section*{References}


\begin{thebibliography}{10}
\expandafter\ifx\csname url\endcsname\relax
  \def\url#1{{\tt #1}}\fi
\expandafter\ifx\csname urlprefix\endcsname\relax\def\urlprefix{URL }\fi
\providecommand{\eprint}[2][]{\url{#2}}

\bibitem{martinez2015permanent}
Martinez E, Daidone A, Gutierrez C, Pera J, Boladeras A, Ferrer F, Pino F,
  Suarez J~F, Polo A and Guedea F 2015 {\em Brachytherapy\/} {\bf 14} 166--172

\bibitem{tanaka2015urethral}
Tanaka N, Asakawa I, Hasegawa M and Fujimoto K 2015 {\em Brachytherapy\/} {\bf
  14} 131--135

\bibitem{horiot1997accelerated}
Horiot J~C, Bontemps P, Van~den Bogaert W, Le~Fur R, van~den Weijngaert D,
  Bolla M, Bernier J, Lusinchi A, Stuschke M, Lopez-Torrecilla J {\em et~al.\/}
  1997 {\em Radiotherapy and Oncology\/} {\bf 44} 111--121

\bibitem{Lyman1985}
Lyman J~T 1985 {\em Radiation Research\/} {\bf 104} S13--S19

\bibitem{Niemierko1993}
Niemierko A and Goitein M 1993 {\em International Journal of Radiation
  Oncology, Biology, Physics\/} {\bf 25} 135--145

\bibitem{hanin2013mechanistic}
Hanin L and Zaider M 2013 {\em Physics in Medicine and Biology\/} {\bf 58} 825

\bibitem{zaider2000tumour}
Zaider M and Minerbo G 2000 {\em Physics in Medicine and Biology\/} {\bf 45}
  279

\bibitem{dawson2006derivation}
Dawson A and Hillen T 2006 {\em Computational and Mathematical Methods in
  Medicine\/} {\bf 7} 121--141

\bibitem{maler2009cell}
Maler A and Lutscher F 2009 {\em Mathematical Medicine and Biology: A Journal
  of the IMA\/} {\bf 27} 313--342

\bibitem{hillen2010cell}
Hillen T, De~Vries G, Gong J and Finlay C 2010 {\em Acta Oncologica\/} {\bf 49}
  1315--1323

\bibitem{stocks2016stochastic}
Stocks T, Hillen T, Gong J and Burger M 2017 {\em Mathematical Medicine and
  Biology: A Journal of the IMA\/} {\bf 34} 469--492

\bibitem{stavrev2001generalization}
Stavrev P, Stavreva N, Niemierko A and Goitein M 2001 {\em Physics in Medicine
  and Biology\/} {\bf 46} 1501

\bibitem{tucker2006cluster}
Tucker S~L, Zhang M, Dong L, Mohan R, Kuban D and Thames H~D 2006 {\em
  International Journal of Radiation Oncology - Biology - Physics\/} {\bf 64}
  1255--1264

\bibitem{Rutkowska2010}
Rutkowska E, Baker C and Nahum A 2010 {\em Physics in Medicine and Biology\/}
  {\bf 55} 2121

\bibitem{dandrea2016}
D'Andrea M, Benassi M~B and Strigari L 2016 {\em Computational and Mathematical
  Methods in Medicine\/} {\bf 2016} 2796186

\bibitem{hendry1986tissue}
Hendry J and Thames H 1986 {\em The British Journal of Radiology\/} {\bf 59}
  628--630

\bibitem{konings2005mechanism}
Konings A~W, Coppes R~P and Vissink A 2005 {\em International Journal of
  Radiation Oncology• Biology• Physics\/} {\bf 62} 1187--1194

\bibitem{dale2007radiobiological}
Dale R~G and Jones B~E 2007 {\em Radiobiological modelling in radiation
  oncology\/} (British Institute of Radiology, London)

\bibitem{bond1965mammalian}
Bond V~P, Fliedner T~M and Archambeau J~O 1965 {\em Mammalian radiation
  lethality: a disturbance in cellular kinetics\/} (Academic Press)

\bibitem{redner2001guide}
Redner S 2001 {\em A guide to first-passage processes\/} (Cambridge University
  Press)

\bibitem{gillespie1976general}
Gillespie D~T 1976 {\em Journal of Computational Physics\/} {\bf 22} 403--434

\bibitem{gillespie1977exact}
Gillespie D~T 1977 {\em The journal of Physical Chemistry\/} {\bf 81}
  2340--2361

\bibitem{gardiner1985handbook}
Gardiner C~W 2004 {\em Handbook of Stochastic Methods\/} (Springer-Verlag,
  Berlin)

\bibitem{gong2011}
Gong J, Dos~Santos M~M, Finlay C and Hillen T 2013 {\em Mathematical Medicine
  and Biology: A Journal of the IMA\/} {\bf 30} 1--19

\bibitem{van1992stochastic}
van Kampen N~G 2007 {\em Stochastic Processes in Physics and Chemistry\/}
  (North-Holland, Amsterdam)

\bibitem{kloeden1992sde}
Kloeden P~E and Platen E 1992 {\em Numerical Solution of Stochastic
  Differential Equations\/} (Springer-Verlag Berlin Heidelberg)

\bibitem{risken1984fokker}
Risken H 1989 {\em The Fokker--Planck Equation: Methods of Solution and
  Applications\/} (Springer-Verlag, Berlin)

\bibitem{ricciardi1988first}
Ricciardi L~M and Sato S 1988 {\em Journal of Applied Probability\/} {\bf 25}
  43--57

\bibitem{madec2004first}
Madec Y and Japhet C 2004 {\em Mathematical Biosciences\/} {\bf 189} 131--140

\bibitem{lo2006computing}
Lo C~F and Hui C~H 2006 {\em Applied Mathematics Letters\/} {\bf 19} 1399--1405

\bibitem{hall2006radiobiology}
Hall E~J and Giaccia A~J 2006 {\em Radiobiology for the Radiologist\/}
  (Lippincott Williams \& Wilkins)

\bibitem{ponce2017stochastic}
Ponce~Bobadilla A~V, Maini P~K and Byrne H 2017 {\em Mathematical Medicine and
  Biology: A Journal of the IMA\/}  dqw024

\bibitem{brenner1997use}
Brenner D~J and Herbert D~E 1997 {\em Medical Physics\/} {\bf 24} 1245--1248

\bibitem{brenner2008linear}
Brenner D~J 2008 The linear-quadratic model is an appropriate methodology for
  determining isoeffective doses at large doses per fraction {\em Seminars in
  radiation oncology\/} vol~18 (Elsevier) pp 234--239

\bibitem{fowler201021}
Fowler J~F 2010 {\em The British Journal of Radiology\/} {\bf 83} 554--568

\bibitem{gibbs2002choosing}
Gibbs A~L and Su F~E 2002 {\em International Statistical Review\/} {\bf 70}
  419--435

\bibitem{kullback1951information}
Kullback S and Leibler R~A 1951 {\em The Annals of Mathematical Statistics\/}
  {\bf 22} 79--86

\bibitem{mcaneney2007investigation}
McAneney H and O'Rourke S 2007 {\em Physics in Medicine \& Biology\/} {\bf 52}
  1039

\bibitem{mundt2005intensity}
Mundt A~J and Roeske J~C 2005 {\em Intensity modulated radiation therapy: a
  clinical perspective\/} vol~1 (People's Medical Publishing House, USA)

\bibitem{kallman1992algorithm}
Kallman P, Lind B~K and Brahme A 1992 {\em Physics in Medicine \& Biology\/}
  {\bf 37} 871

\bibitem{carlson2004comparison}
Carlson D~J, Stewart R~D, Li X~A, Jennings K, Wang J~Z and Guerrero M 2004 {\em
  Physics in Medicine and Biology\/} {\bf 49} 4477

\bibitem{kishan2015late}
Kishan A~U and Kupelian P~A 2015 {\em Brachytherapy\/} {\bf 14} 148--159

\bibitem{metzler2014}
Metzler R, Oshanin G and Redner S~E 2014 {\em First-Passage Phenomena and Their
  Applications\/} (World Scientific, Singapore)

\bibitem{lea1942mechanism}
Lea D and Catcheside D 1942 {\em Journal of Genetics\/} {\bf 44} 216--245

\end{thebibliography}

 \providecommand{\newblock}{}

\begin{appendix}

\section{The LQ formalisation}\label{app:LQ}
We briefly review the LQ formalism for a radioactive implant \cite{brenner1997use,brenner2008linear,fowler201021}. We first consider the reaction describing death due to irradiation. The LQ formalism relates the mean surviving fraction of cells $\psi$ to the total dose delivered in a time interval $\left[0, t\right]$, $D(t)$: 
\begin{equation}
\psi(t)=e^{-\alpha D(t) - \beta q(t) D(t)^2 }.
\label{eq:psi}
\end{equation} 
Here, there are two radiosensitivity parameters, $\alpha$ and $\beta$, which describe a tissue's linear and quadratic responses to a source of radiation, respectively. For a radioactive source exponentially decaying with rate $\lambda$ and with an initial dose rate $R_0$, the total dose delivered by time $t$ is given by $D(t)=R_0/\lambda\left[1-\exp(-\lambda t)\right]$.
The function $q(t)$ in Eq.~\eqref{eq:psi} is the Lea--Catcheside protraction factor \cite{lea1942mechanism}, which is specific to the method of treatment involved. In the case of brachytherapy it is given by
\begin{equation}
q(t)=\frac{2(\lambda t)^2}{(\gamma t)^2 (1-\lambda^2 / \gamma^2) \left(1-e^{-\lambda t} \right)^2} 
\left[e^{-(\lambda+\gamma)t} + \gamma t \left( \frac{1 - e^{-2 \lambda t}}{2 \lambda t} \right) - \frac{1 + e^{-2 \lambda t}}{2}\right].
\end{equation}
Here, $\gamma$ is the rate at which radiation-damaged cells repair their DNA.
The fractional change in the population over an infinitesimal time $\dot\psi(t)/\psi(t)$ gives the hazard function $h(t)$. This is found to be given by \cite{stocks2016stochastic}
\begin{equation}
h(t)=\alpha R_0 e^{-\lambda t} + \frac{2 \beta R_0^2 e^{-\lambda t} }{\gamma - \lambda} \left( e^{-\lambda t} - e^{- \gamma t}\right).
\label{eq:h_brachy}
\end{equation}

\section{Evaluation of Approximation $1$ for the model of normal and doomed cells in Sec.~\ref{ssec:model2d}}
\label{sec:2d_calculation}
We write $N_t$ for the number of normal cells at time $t$ and $X_t$ for the number of doomed cells.
We are interested in the population of total functional cells, $S_t\equiv N_t+X_t$. Specifically, we are interested in the time $S_t$ first passes a boundary $L$. The master equation can be formulated in terms of $S$ and $X$:
\begin{align}\begin{split}
\frac{\d}{\d t}P_{S,X}(t)={}
 &\left(\E_S^{-1}-1\right)b_0 (S-X)\left(1-\frac{S}{k M}\right) P_{S,X}(t)\\
&+\left(\E_X^{-1}-1\right) h(t) (S-X) P_{S,X}(t)\\
&+\left(\E_S^{+1}-1\right) d_1 (S-X) P_{S,X}(t)\\
&+\left(\E_S^{+1}\E_X^{+1}-1\right) d_2 X P_{S,X}(t),
\end{split}\end{align}
where $P_{S,X}(t)$ is the probability that random processes $S_t$, $X_t$ have the values $S$, $X$ at time $t$. The operator $\E_S$ is the step operator affecting the size of the total population, and $\E_X$ is the step operator affecting the number of doomed cells, i.e. $\E_S f_{S,X}=f_{S+1,X}$ and $\E_X f_{S,X}=f_{S,X+1}$.

We proceed by approximating the master equation via a Kramers--Moyal expansion. First, we introduce re-scaled processes $s_t=S_t/M$ and $x_t=X_t/M$, and then expand the step operators in the limit $M\gg 1$. We arrive at the Fokker--Planck equation
\begin{align}\begin{split}
\frac{\partial}{\partial t}\Pi(s,x,t)={}
&-\frac{\partial}{\partial s}\left[ b_0\left(1-\tfrac{s}{k}\right)(s-x) - d_1 (s-x) - d_2 x \right] \Pi(s,x,t)\\
&-\frac{\partial}{\partial x}\left[ h (t) (s-x) - d_2 x \right] \Pi(s,x,t)\\
&+\frac{1}{2M}\frac{\partial^2}{\partial s^2}\left[ b_0\left(1-s\right)(s-x) + d_1(s-x) + d_2 x \right] \Pi(s,x,t)\\
&+\frac{1}{2M}{\frac{\partial^2}{\partial x^2}}\left[ h (t) (s-x)+ d_2 x \right] \Pi(s,x,t)\\
&+\frac{1}{M}\frac{\partial}{\partial s}\frac{\partial}{\partial x}~d_2 x \Pi(s,x,t),
\end{split}\end{align}
where we have neglected higher-order terms in $M^{-1}$. This Fokker--Planck equation can equivalently be written as an SDE:
\begin{align}\label{eq:sxSDE}
\left(\begin{matrix}
\d s_t\\
\d x_t
\end{matrix}\right)=
\vec{\mu}(s_t,x_t)\d t
+\frac{1}{M^{1/2}}\textbf{B}(s,x,t)
\left(\begin{matrix}
\d W_t^{(1)} \\
\d W_t^{(2)}
\end{matrix}\right),
\end{align}
where the drift is given by
\begin{align}
\vec{\mu}(s,x)=\left(\begin{matrix}
b\left(1-\frac{s}{k}\right)(s-x) - d_1 (s-x) - d_2 x\\
h(t) (s-x) - d_2 x
\end{matrix}\right).\
\end{align}
The diffusion $\textbf{B}(s,x,t)$ is the positive-semidefinite matrix satisfying
\begin{align}
\textbf{B}^2(s,x,t)=
\left(\begin{matrix}
b\left(1-\frac{s}{k}\right)(s-x) + d_1 (s-x) + d_2 x & d_2 x\\
d_2 x & h(t) (s-x) + d_2 x
\end{matrix}\right).
\end{align}
We proceed by linearising the SDE~\eqref{eq:sxSDE}. Let $s_t=\phi_1(t)+M^{-1/2}{\xi_1}_t$ and $x_t=\phi_2(t)+M^{-1/2}{\xi_2}_t$, where $\phi_1(t)$ and $\phi_2(t)$ are the deterministic functions of time. Substituting and collecting lowest order terms, we see these functions are given by the ODEs
\begin{subequations}\begin{align}
\frac{\d \phi_1}{\d t} ={}& \left(1-\frac{\phi_1}{k}\right) b(\phi_1-\phi_2) - d_1 (\phi_1-\phi_2 ) - d_2 \phi_2, \label{eq:help}\\
\frac{\d \phi_2}{\d t} ={}& h (t)(\phi_1-\phi_2) - d_2 \phi_2,
\end{align}\label{eq:2d_ODES}\end{subequations}
i.e., we recover Eqs.~\eqref{eq:det2d}. 

The random processes ${\xi_1}_t$ and ${\xi_2}_t$ describe deviations from this deterministic trajectory, and are of the Ornstein--Uhlenbeck type
\begin{align}\label{eq:doomedLNA}
\d \vec{\xi}_t=
\textbf{ A}(\phi_1,\phi_2,t) \, \vec{\xi}_t 
\d t
+\textbf{ B}(\phi_1,\phi_2,t) \, \d \vec{W}_t,
\end{align}
where $\textbf{A}(\phi_1,\phi_2,t)$ is given by
\begin{align}
\textbf{A}(\phi_1,\phi_2,t) =-\left(\begin{matrix}
b\left(1-2\frac{\phi_1}{k}+\frac{\phi_2}{k}\right)-d_1 & b\left(\frac{\phi_1}{k}-1\right) + d_1 - d_2\\
h(t) & -h(t)-d_2
\end{matrix}\right).
\end{align}
We note that the argument of $\textbf{B}$ in Eq.~\eqref{eq:doomedLNA} is now given by $\phi_1$ and $\phi_2$, so that the noise is additive rather than multiplicative. 

We are interested in the variation of the total population size from the deterministic path $\left<{\xi_1}_t^2\right>$; we remark that by construction $\left<{\xi_1}_t\right>=\left<{\xi_2}_t\right>=0$. The variances and covariance of ${\xi_1}_t$ and ${\xi_2}_t$ can be seen to evolve in time as follows \cite{risken1984fokker}
\begin{subequations}\begin{align}
\frac{\d \left< {\xi_1}_t^2 \right>}{\d t} ={}& 2 A_{11} \left< {\xi_1}_t^2 \right> + 2 A_{12} \left< {\xi_1}_t {\xi_2}_t \right> + (B_{11})^2 + (B_{12})^2,\\
\frac{\d \left< {\xi_2}_t^2 \right>}{\d t} ={}& 2 A_{22} \left< {\xi_2}_t^2 \right> + 2 A_{21} \left< {\xi_1}_t {\xi_2}_t \right> + (B_{22})^2 + (B_{21})^2,\\
\frac{\d \left< {\xi_1}_t {\xi_2}_t \right>}{\d t} ={}& A_{21}\left< {\xi_1}_t^2 \right>+A_{12}\left< {\xi_2}_t^2 \right> +\left(A_{11}+A_{22}\right)\left< {\xi_1}_t {\xi_2}_t \right> + B_{11}B_{21} + B_{12}B_{22}.
\end{align}\label{eq:noise_ODEs}\end{subequations}
For a given set of parameters, we numerically integrate the five coupled Eqs.~\eqref{eq:2d_ODES} and Eqs.~\eqref{eq:noise_ODEs}. This provides the mean and covariance matrix for the bivariate Gaussian distribution of the number of normal and doomed cells as a function of time. For Approximation $1$, the time $t^*$ is defined by $\phi_1(t^*)=\ell$; this is the point in time when the total number of functional cells crosses the threshold for onset of NTC. The variance of the number of functional cells at this time is given by $\Sigma^2(t^*)=\left<{\xi_1}_{t^*}^2\right>$ within the LNA. We then use Eq.~\eqref{eq:erf}, where $\mu(\ell,t^*)$ is to be replaced by the right-hand side of Eq.~(\ref{eq:help}), evaluated at $t^*$.

Approximation $2$ is computed using Eq.~\eqref{eq:q}, replacing $\phi(t)$ by $\phi_1(t)$, and $\Sigma^2(t)$ by $\left<{\xi_1}_{t}^2\right>$, respectively.

\end{appendix}
 
\end{document}